\documentclass{aa}  
\usepackage{graphicx}

\newcommand{\HI}{{\ion{H}{1}}}

\newcommand{\kms}{$\,$km$\,$s$^{-1}$}

\newcommand{\ergs}{$\,$erg$\,$s$^{-1}$}

\newcommand{\mJybeam}{mJy beam$^{-1}$}

\newcommand{\msun}{{$M_\odot$}}
\newcommand{\msunyr}{{$M_\odot$ yr$^{-1}$}}

\newcommand{\pks}{{PKS\,0023$-$26}}
\newcommand{\coOne}{{CO(1-0)}}
\newcommand{\coTwo}{{CO(2-1)}}

\def\OIII{{[\ion{O}{III}]}}
\def\HI{\ion{H}{i}}

\def\emph#1{{\sl #1}}
\newcommand{\ltsima} {$\; \buildrel < \over \sim \;$}
\newcommand{\gtsima} {$\; \buildrel > \over \sim \;$}
\newcommand{\lta} {\lower.5ex\hbox{\ltsima}}
\newcommand{\gta} {\lower.5ex\hbox{\gtsima}}

\usepackage{txfonts}
\begin{document} 

   \title{Taking snapshots of the jet-ISM interplay: The case of PKS~0023--26} 

\authorrunning{Morganti
 et al.}
\titlerunning{PKS~0023--26 with ALMA}
\author{Raffaella Morganti\inst{1,2}, Tom Oosterloo\inst{1,2}, Clive Tadhunter\inst{3}, Emmanuel P. Bernhard\inst{3}, J. B. Raymond Oonk\inst{4,1,5} }
\institute{ASTRON, the Netherlands Institute for Radio Astronomy, Oude Hoogeveensedijk 4, 7991 PD, Dwingeloo, The Netherlands. 
% \email{morganti@astron.nl}
\and
Kapteyn Astronomical Institute, University of Groningen, Postbus 800,
9700 AV Groningen, The Netherlands
\and
Department of Physics and Astronomy, University of Sheffield, Sheffield, S7 3RH, UK
\and
SURFsara, Postbus 94613, 1090 GP Amsterdam, The Netherlands
\and
Leiden Observatory, Leiden University, Postbus 9513, 2300 RA Leiden
}
 \abstract
 {
We present high  angular resolution (0.13 -- 0.4 arcsec) ALMA \coTwo\ and 1.7 mm continuum observations of the far-infrared-bright galaxy \pks\  ($z = 0.32$), which hosts a young radio source as well as a   luminous optical active galactic nucleus (AGN). Although young, the powerful radio 
source has already grown to a size of a few kiloparsec, making it potentially capable of affecting the  interstellar medium (ISM) of the host galaxy.  

We detect a  very extended distribution of molecular gas with a mass between 0.3 and 3 $\times 10^{10}$ \msun, depending on the $X_{\rm CO}$ conversion factor. The gas has a maximum radial extent of  $\sim$5 arcsec (24 kpc) from the nucleus and is distributed in an asymmetric structure offset from the radio galaxy and with a fairly smooth velocity gradient. At large radii,  tails of gas are observed in the direction of companion galaxies, suggesting that tidal interactions may be responsible for the origin of the gas.  Overall, the observed properties are reminiscent of the molecular structures observed in some galaxy clusters.
However, in the inner few kiloparsec,  across the entire extent of the radio  continuum, the kinematics of the gas appears to be affected by the radio source. 
In the central, sub-kiloparsec region, we observe the brightest emission from the  molecular gas and the broadest velocity profiles with a full width at zero intensity (FWZI) of $\sim$500 \kms, which indicate that in this region  a direct interaction of the jet with dense clouds and outflowing molecular gas is happening. 
On larger, kiloparsec-scales, the molecular gas appears to avoid the radio lobes, while gas with a somewhat smaller velocity dispersion (FWZI of $\sim$350\kms) is observed around the radio lobes. Thus, in these regions, the gas appears to be affected by the  expanding cocoon surrounding the  radio source, likely dispersing and heating preexisting molecular clouds. 
The observations suggest that  the mode of coupling between radio jets and the ISM  changes from  an outflowing phase limited to the sub-kiloparsec region to a maintenance phase, excavating cavities devoid of dense gas,  at larger radii. This reveals that, already on galaxy scales, the impact of the AGN is  not limited to outflows. This is in accordance with predictions from numerical simulations.

With a star-formation rate of  25 \msunyr, \pks\ is located on the SFR-$M_*$ relation for star forming galaxies. Thus, the AGN does not appear to have, at present, a major impact on the host galaxy in terms of the overall level of star-formation activity. However, as the jet and lobes expand throughout the galaxy in the coming  few $\times 10^7$ yr, they  will carry  enough energy to  be  able  to  prevent  further  gas  cooling and/or to inject turbulence and thus affect future star formation. 
 }
   \keywords{galaxies: active - galaxies: individual: PKS~0023--26 - ISM: jets and outflow - radio lines: galaxies}
   \maketitle  

%-------------------------------------------------------------------

\section{Introduction}
\label{sec:introduction}

The evolution of massive galaxies is considered to be strongly influenced by the energy released by their active super massive black holes (SMBHs). This process - also known as feedback - depends on a variety of parameters that observations and theory are now starting to unravel. 
In the commonly assumed picture of feedback by active galactic nuclei (AGN), radio jets are considered to be mostly responsible  for the so-called ``maintenance mode'' (see, e.g., \citealt{McNamara12}). In this mode, the energy associated with momentum-driven lobes of radio AGN acts to prevent the cooling of the gas at large scales, in the warm and hot circumgalactic medium (CGM; or halo) and the hot intergalactic medium (IGM). This mode complements the radiation-driven outflows and winds considered to dominate in the ``quasar mode''  and responsible for clearing the galaxy from gas  at smaller radii. 

Observational signatures of AGN impact are reported in a large number of studies (see \citealt{McNamara12,King15,Harrison18,Veilleux20} for some recent reviews) confirming the relevance of this form of feedback. However, with more detailed observations becoming available (e.g.,  tracing multiple phases of the gas in a large variety of AGN), the picture of AGN feedback has become more complex than the two modes of feedback described above (see, e.g., \citealt{Harrison21,Scholtz20}). At the same time, cosmological simulations, becoming increasingly more sophisticated, require,  for a more realistic implementation, better constraints on the processes of transferring the energy from the active nucleus to the interstellar medium (ISM) and on the location where this happens (see, e.g.,  \citealt{Schaye15,Weinberger17,Dave19,Zinger20}).  

Gaseous outflows have been found with a high incidence in AGN  \citep{Tadhunter08,Harrison18,Veilleux20}, as expected from the  feedback scenario. However, their observed impact - in terms of, for example, mass outflow rate, what part of the galaxy is affected, and the amount of gas escaping the galaxy - is not always as large as predicted by cosmological simulations. Outflows are often limited to the central kiloparsec or sub-kiloparsec regions of the galaxy, and only relatively small amounts of gas seem to be involved (see, e.g.,  \citealt{Holt08,Rose18,Oosterloo17,Oosterloo19,Baron19,Bischetti19,Scholtz20,Santoro18,Santoro20}).
Thus, the impact of an active SMBH may be sublter and, in particular, may even change during the various stages in the life of an AGN.
In order to learn more about this, here we investigate the impact of young radio jets by spatially tracing the properties of the molecular gas around the radio source of \pks, and we use the results to understand whether the way AGN feedback occurs (and more in general the coupling between radio plasma and the ISM) depends on the evolutionary stage of the jet.  

An increasing number of observations indicate that radio jets can impact the ISM  on galaxy scales (for some examples and reviews see  \citealt{Nesvadba08,Holt08,Holt09,Morganti18,Hardcastle20}). Outflows driven by jets have been observed in a growing number of cases. Interestingly, this could be the case not only in objects with jets of high radio power, but also in the more common low radio power sources (e.g.,  \citealt{Rodriguez17,Fabbiano18,Maksym19,Murthy19,Husemann19a,Jarvis19,Audibert19}).

Newly born radio jets,  observed in compact steep-spectrum  (CSS) and giga-hertz peaked(GPS) sources, appear to be particularly effective in driving gas outflows (see, e.g.,  \citealt{Holt08,Holt09,Morganti05a,Shih13,Aditya18,Molyneux19,Morganti18} and refs therein).  
However, whether, and how, this impact evolves as the jet grows is less clear.
For example, the largest impact of outflows may be limited in time because the radio source quickly grows beyond the  denser ISM in the central kiloparsec region \citep{Gereb15,Oosterloo19}.  
Such an evolution in the effects of the radio jets has been predicted by  numerical simulations. In particular, the work of \cite{Sutherland07} has suggested the presence of four phases in the evolution of the jets, each providing a different type of impact on the surrounding ISM: 1) an initial ‘‘flood and channel’’ phase, where the expansion of the jet strongly depends on the interaction with high-pressure gas near the AGN;  2) following this, and as consequence of this interaction, the formation of a spherical, energy-driven bubble phase;  3) a subsequent, rapid phase  where the jet breaks free from the last obstructing dense clouds in the ISM and, finally, 4) a classical phase, where the jet propagates to large scales into the CGM and IGM in a momentum-dominated fashion.

In order to trace these different phases and how the impact of the jets may change with this evolution, we need to explore a large parameter space. This requires observing radio galaxies in different phases of their evolution and tracing the impact of their jets using different phases of the gas. 
The object presented here, \pks, was selected to complement the work done so far by our group, which has been performing detailed studies of young radio galaxies using cold (molecular and \HI) and warm ionised gas to cover the parameter space mentioned above  \citep[][and references therin]{Morganti15,Dasyra16,Oosterloo17,Maccagni18,Murthy19,Oosterloo19,Schulz18,Schulz21}.
The radio emission in \pks\ extends to few kiloparsec, thus probing interesting intermediate scales and providing a bridge between feedback from young sub-kiloparsec radio jets and what happens on larger galactic scales in more evolved sources.

%-------------------------------------- Environment
   \begin{figure}
   \centering
\includegraphics[angle=0,width=8cm]{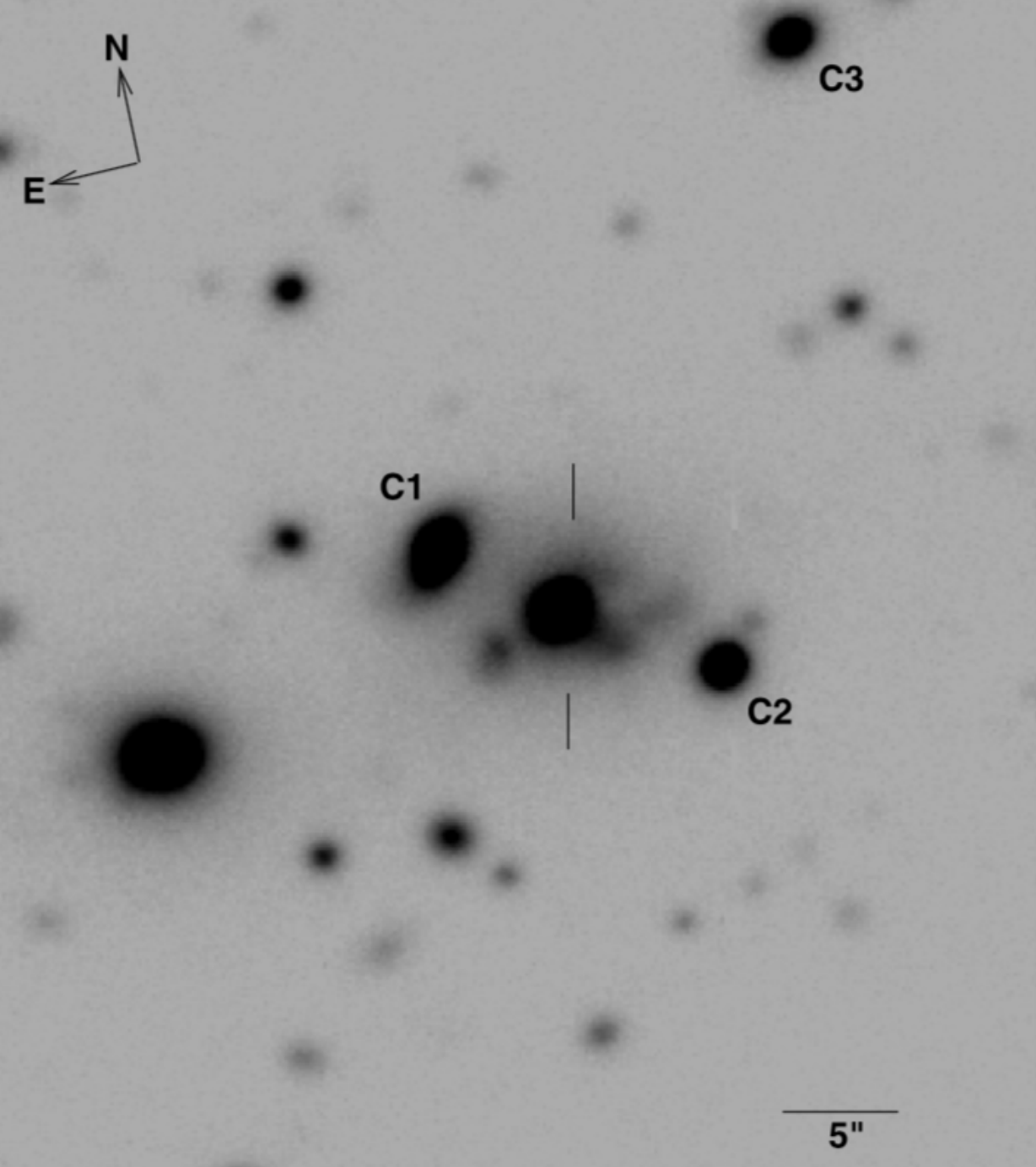}
   \caption{
   Optical image from Gemini \citep{Ramos13}. Marked as C1, C2, and C3 are galaxies confirmed to have redshifts similar to that of \pks. More objects and tails are seen even closer (projected distances $\sim$10 kpc) to the target galaxy.
  }
              \label{fig:environment}
    \end{figure}
%--------------------------------------

Here we present new ALMA CO(2-1) observations obtained with sufficient angular resolution  ($\sim$0\farcs4) to be able to trace the properties and kinematics of the molecular gas spatially associated with the radio-emitting plasma, which has an extent of 
about 1 arcsec,  and the interplay between the two.
We describe the known properties of \pks\ in Sect.\  \ref{sec:description0023} and the new ALMA observations in Sect.\  \ref{sec:observations}. The results of the \coTwo\ and the millimeter continuum emission are given in Sects. \ref{sec:molecular} and  \ref{sec:continuum}, respectively. A possible scenario and the implications for the evolution of the host galaxy of \pks\ are given in Sect.\  \ref{sec:scenario}.

\section{An overview of the properties of \pks}
\label{sec:description0023}

\pks\ is a young, powerful radio source ($\log P_{\rm 5\, GHz}/{\rm W\, Hz^{-1}} = 27.43$) hosted by an early-type galaxy (see \citealt{Ramos11} and refs therein).
An accurate redshift of $z = 0.32188(4)$ has been derived by \cite{Santoro20} by fitting stellar absorption lines in an X-shooter spectrum, refining the estimate of  \cite{Holt08}\footnote{At this redshift, 1 arcsec corresponds to 4.716 kpc for the cosmology adopted in this paper which assumes a flat Universe and the following parameters: $H_{\circ} = 70$ \kms\ Mpc$^{-1}$, $\Omega_\Lambda = 0.7$, $\Omega_{\rm M} = 0.3$.}. The radio continuum emission is distributed in two relatively symmetric lobes as described in \cite{Tzioumis02}. The extent of the radio emission is about 3 kpc measured from peak to peak of the lobes. The full extent could be larger (up to $\sim 4.7$ kpc), since the lobes may include some  diffuse emission at larger radii that is 
tentatively detected in the MERLIN image presented in \cite{Tzioumis02}, as well as in the images presented in this paper.

\begin{table*}
\caption{Parameters of the ALMA data cubes and millimeter images. 
}
\begin{center}
\begin{tabular}{lcr@{ $\times$ }lccccc} 
\hline\hline 
%\multicolumn{3}{c}{WSRT}    \\
   & Frequency & \multicolumn{3}{c}{Beam \& PA} &Velocity res.  & Noise     & Description\\
   &  (GHz)   & \multicolumn{2}{c}{(arcsec) }  & (degree)& (\kms)          & (\mJybeam) & \\
\hline
\coTwo\   &        &  0.45&0.34 & --77.8  & 34   & 0.117   & low spatial resolution, robust = +2  \\
%\coTwo\   &        &  0.21&0.17 & --79.1  & 34    & 0.123 & intermediate resolution \opm{where do we use this?} \\
\coTwo\   &        &  0.15&0.11 & --80.9  & 1.7   & 0.4  &  CO absorption, robust = --2  \\
Continuum &  176   &  0.45&0.34 & --77.8  &  --   & 0.08 &  robust = +2 \\
%Continuum &  176   &  0.21&0.17 & --79.1  &  --   & 0.15 &  \opm{weighting  rob=0.5?}\\
Continuum &  176   &  0.15&0.11 & --80.9  &  --   & 0.16 &  robust = --2\\
\hline
\end{tabular}
\tablefoot{ The maximum recoverable scale of the \coTwo\ observations is 3.8 arcsec.}
\end{center}
\label{tab:obs}
\end{table*}
Strong optical emission lines have been detected by \citet{Holt08}, \citet{Shih13} and \citet{Santoro20}. The \OIII\ luminosity ($1.5 \times 10^{35}$ W; \citealt{Dicken09}) and a bolometric luminosity in the range $2.5 - 4 \times 10^{45}$ \ergs\  \citep{Santoro20}  make \pks\ just qualify as a type 2 quasar  \citep{Holt08} according to \citet{Zakamska06}.
Warm ionised gas is seen in the central 10 kpc \citep{Shih13}. Two separate components of the emission lines are observed, with the broad one concentrated around the centre, in a region to first order co-spatial with the radio emission. The presence of disturbed kinematics is indicated by the large width of the broad component (with \OIII\ showing a full-width half maximum (FWHM) width of $\sim$1000 \kms; \citealt{Holt08,Shih13}). 
The integral field  data presented by \cite{Shih13} show that the distribution of the narrower component is instead more elongated along an axis with position angle $\simeq -44^\circ$,  which is slightly misaligned from the radio axis, which in turn has position angle $\simeq -34^\circ$. Despite the clear signatures of interaction  \citep{Holt08,Santoro20}, the ionisation of the gas does not appear to be dominated by a shock component \citep{Santoro20}.  

Neutral atomic hydrogen (\HI) was detected in absorption by \cite{Vermeulen03}, who described the absorption as consisting of two components with FWHM = 39 and 126 \kms\ respectively. Adopting the redshift of \cite{Santoro20}, the deeper and broader component is close to the systemic velocity of \pks, while the second component is redshifted (about 100 \kms). Due to the limited spatial resolution, no information is available about the  location of these components across the source.

The optical continuum displays a large  UV excess at $5 < r < 15$ kpc  (see Fig.\ 1 in \citealt{Holt07}). This, as well as spectral synthesis modelling of the optical spectrum, provides evidence  that the stellar population has a large and spatially extended contribution (50-60\%) from a young stellar population, with an age in the range 0.03 - 0.05 Gyr.  The detection of a far-IR excess and strong mid-IR PAH emission features \citep{Dicken09,Dicken12} is further evidence for strong recent star-formation activity. From modelling the infrared spectral energy distribution (SED), a star-formation rate (SFR) of   $25\pm 2$ \msunyr\ has been derived \citep{Bernhard21}.

Attempts to quantify the large-scale environment of PKS 0023--26  have produced ambiguous results. On the one hand, the angular clustering amplitude derived by \citet{Ramos13} using galaxy number counts provides evidence that \pks\ is in a moderate-density cluster environment, albeit with a high degree of uncertainty. In this context, it is notable that three neighbouring galaxies are confirmed to have similar redshifts \citep{Tadhunter11}, including  (in addition to C3) the two galaxies on either side of \pks\ (at 25~kpc SW and 33~kpc NE, marked as C1 and C2 in Fig.\  \ref{fig:environment}). 
\pks\ and its neighbours also appear to share a common stellar envelop, which has an amorphous appearance close to the radio galaxy, and is possibly crossed by dust lanes \citep{Ramos11}. 

On the other hand, based on XMM observations, the total X-ray luminosity of \pks\ and its surroundings ($\log L_{\rm 2-10\,keV}/{\rm erg s^{-1}} = 43.27$; \citealt{Mingo14})  is lower than expected for a rich cluster of galaxies, and is more consistent with that of a galaxy group \citep{Eckmiller11}. This X-ray luminosity may actually represent an upper limit, given that the emission shows two spatial peaks (separated by $\sim$20 arcsec), one of them centred on the host galaxy of \pks, as well as that the total X-ray luminosity is likely to contain a substantial contribution from the AGN. Upcoming Chandra observations will allow us to clarify this.

\section{ALMA observations of \coTwo\ and 1.7 mm continuum}
\label{sec:observations}

The \coTwo\ and 1.7 mm continuum data  were obtained during Cycle 6 using ALMA in configuration C43-6. The observations were done in four observing sessions (two on Sept 27, 2019 and two on Sept 28, 2019), resulting in a total on-source time of 45 min, with the number of antennas ranging between 45 and 50, and a $uv$ coverage with the shortest baseline being 15 m and a maximum baseline of about 2.6 km.  
The angular resolution of the observations of 0\farcs13 -- 0\farcs4   was chosen to allow the distribution of the molecular gas to be traced, given the sensitivity, on similar spatial scales as the extended radio emission (about 1 arcsec).
The phase centre of the observations was set on the nucleus of \pks, with a field of view (FoV) of $\sim$34”. 

The observations were done in Band 5 making use of the correlator in Frequency Division Mode.  One spectral window was centred on the  \coTwo\ emission using a central frequency 174.436 GHz  and a total bandwidth of 1.875 GHz (corresponding to 5625 \kms). With the 1920 channels used we had a native velocity resolution of 1.7 \kms, but in the subsequent data reduction channels were  averaged to make image cubes with a velocity resolution better matching the observed line widths (see below). 
Three continuum  spectral windows were also used, centred on 176.3, 174.4 and 164.3 GHz, each band having 128 channels and a total width of 2 GHz.
The source J2258--2758 was used as flux, phase and bandpass calibrator  and was  observed for about 10 minutes in total.

The  calibration was done in CASA (v5.1.1; \citealt{McMullin07}) using the  reduction scripts provided by the ALMA observatory. We  assigned a minimum error of 5\% to the flux estimates used \citep{vanKempen14}.
The maximum recoverable scale of the \coTwo\ observations is 3.8 arcsec.
The visibility products (continuum and line) resulting from the ALMA pipeline were found to be of sufficient quality and no further calibration was done. The calibrated $uv$ data were 
exported to MIRIAD \citep{Sault95} to perform further analysis.

\begin{figure}
   \centering
      \includegraphics[angle=0,width=7.5cm]{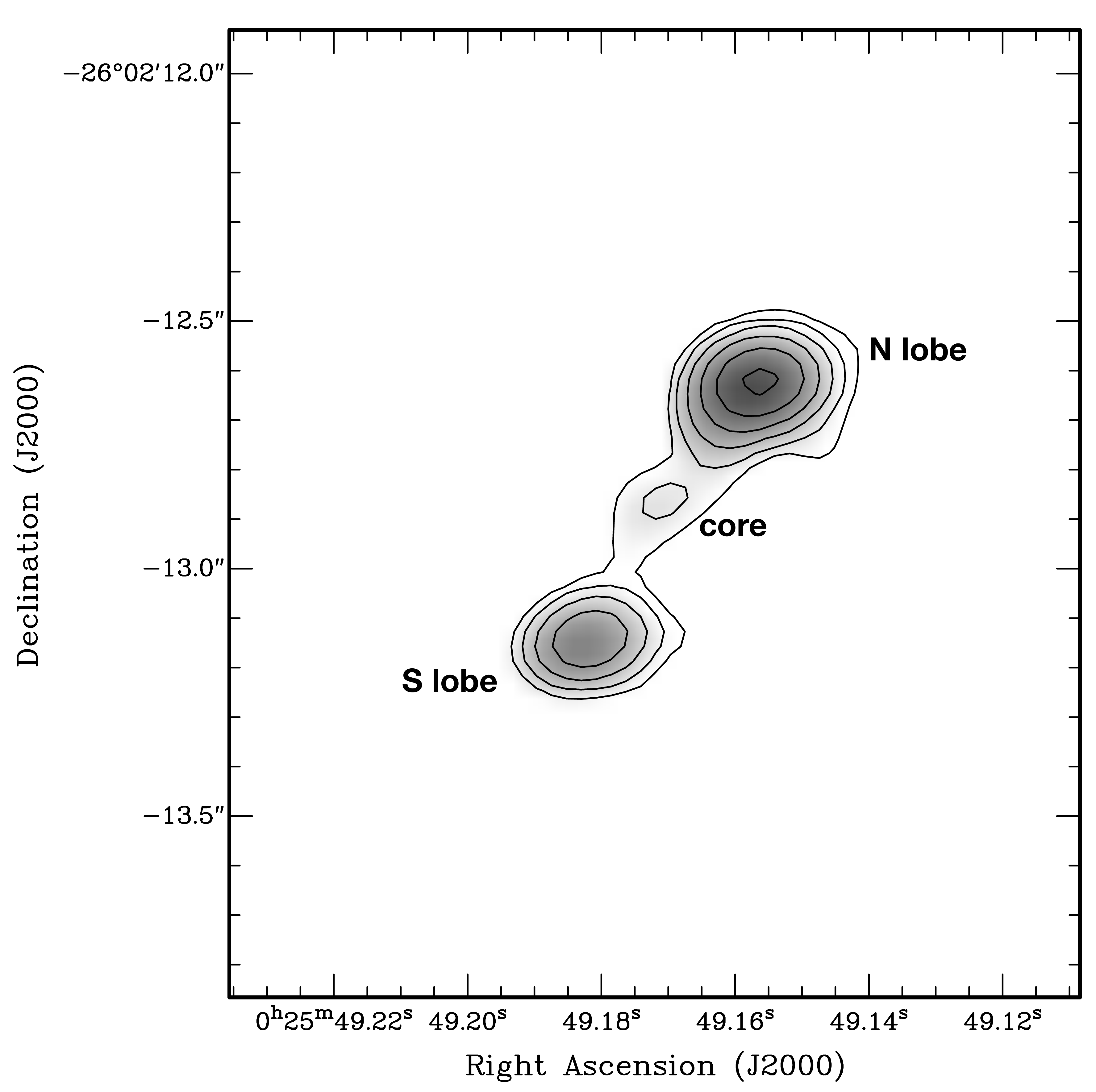}
   \caption{ 1.7 mm continuum emission, derived from the uniformly weighted data. The contours are 1.3, 2.6, 5.2, 10.4, 20.8, and 41.6 \mJybeam. }
              \label{fig:ContUnif}
    \end{figure}

For the line data, two different data cubes were made using different weighting schemes. Their properties are summarised in Table \ref{tab:obs}. 
We made a cube at low spatial resolution using natural weighting (i.e., robust = +2) having  a  beam  $0\farcs45 \times 0\farcs34$, PA $=  -77^\circ$  and a   noise level of 0.12 \mJybeam\ for a velocity resolution of 34 \kms\ in order to optimise the detection of extended and low surface brightness emission. There is a difference between how MIRIAD and CASA estimate the size of the restoring beam. For naturally weighted images having PSF peaks which are strongly non-Gaussian, this leads to a larger (by $\sim$30\%) restoring beam used by MIRIAD for the same data set. 
In addition, a cube at full spatial  and velocity resolution   was obtained using uniform weighting (robust = --2,  see Table \ref{tab:obs}) to characterise the absorption seen against the northern radio lobe (see Sect.\ \ref{sec:absorption}). The continuum emission (Fig.\ \ref{fig:ContUnif}) was subtracted from the data cubes in the image domain by doing, for every position, a linear fit to the line-free channels.

\begin{figure*}
   \centering
    \includegraphics[angle=0,height=5.4cm]{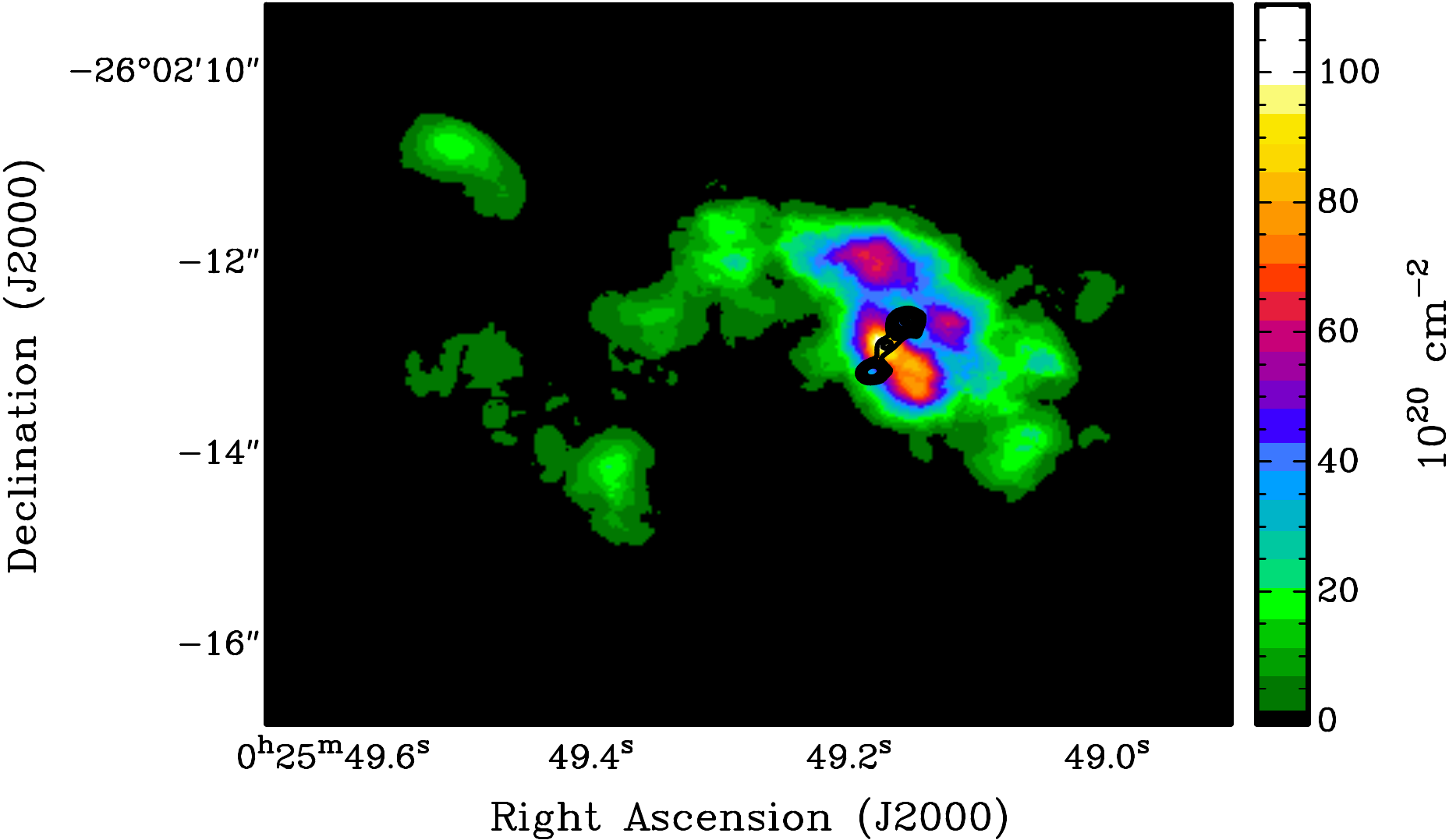}
    \includegraphics[angle=0,height=5.4cm]{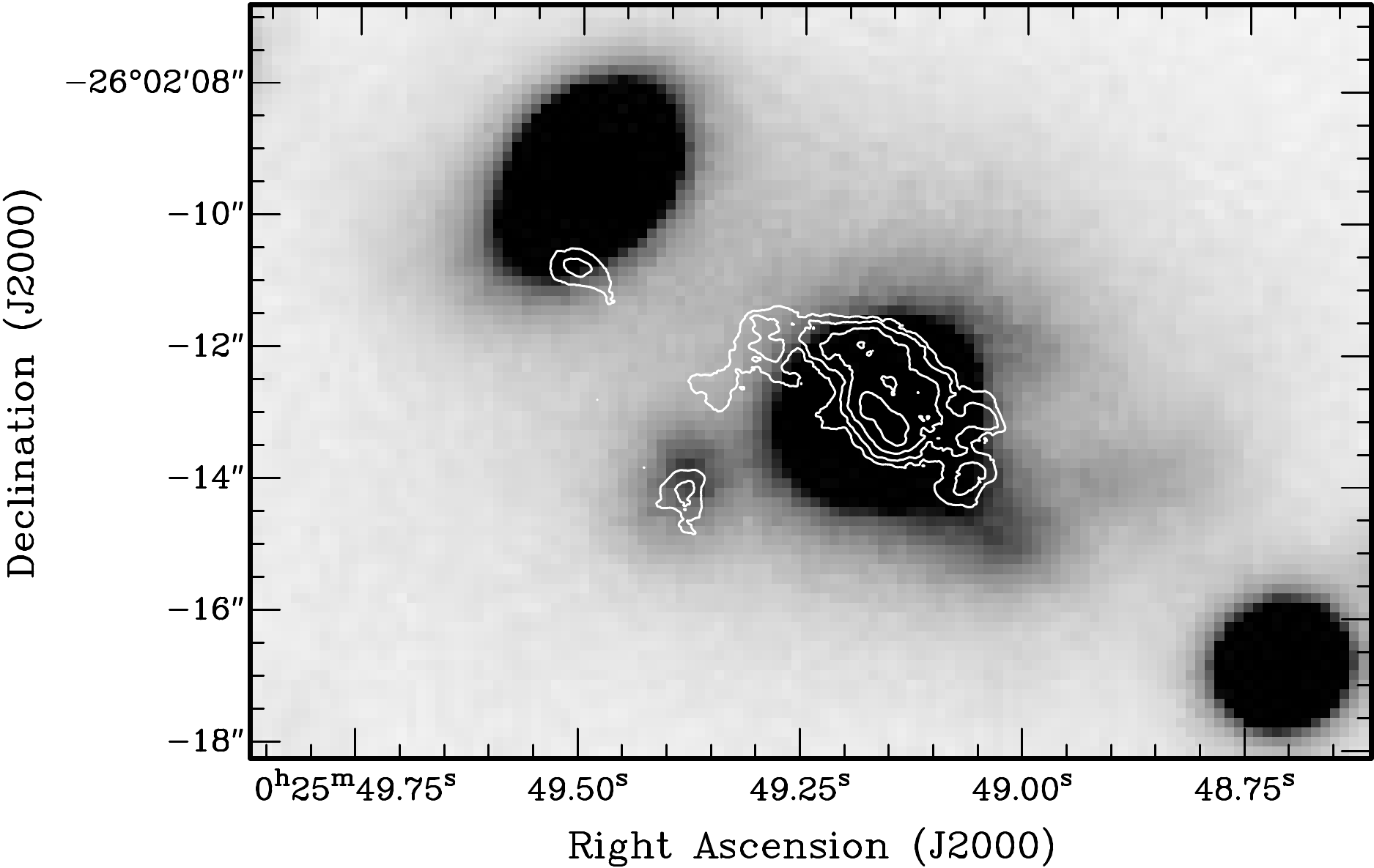}
   \caption{{\bf Left:} Column density image of the molecular gas with superimposed the contours of the continuum emission. The  core of the radio emission is coincident with the peak of the molecular gas (see text for details). Contour levels are 1.3, 2.
6, 5.2, 10.4, 20.8, and 41.6 \mJybeam.  The rms noise of the CO and continuum images can be found in Table \ref{tab:obs} {\bf Right:} Column density contours of the molecular gas  over-plotted to the optical image from \cite{Ramos11}. The offset of the molecular gas compared to the optical galaxy hosting \pks\ is clearly seen. The arm-like structure  to the E points toward one of the close companions where CO emission is also detected. Contour levels are 10, 20, 50, 100, 200, and 500 $\times 10^{20}$ cm$^{-2}$. }
              \label{fig:molecularGas}
    \end{figure*}

\begin{figure*}
   \centering
      \includegraphics[angle=0,width=9cm]{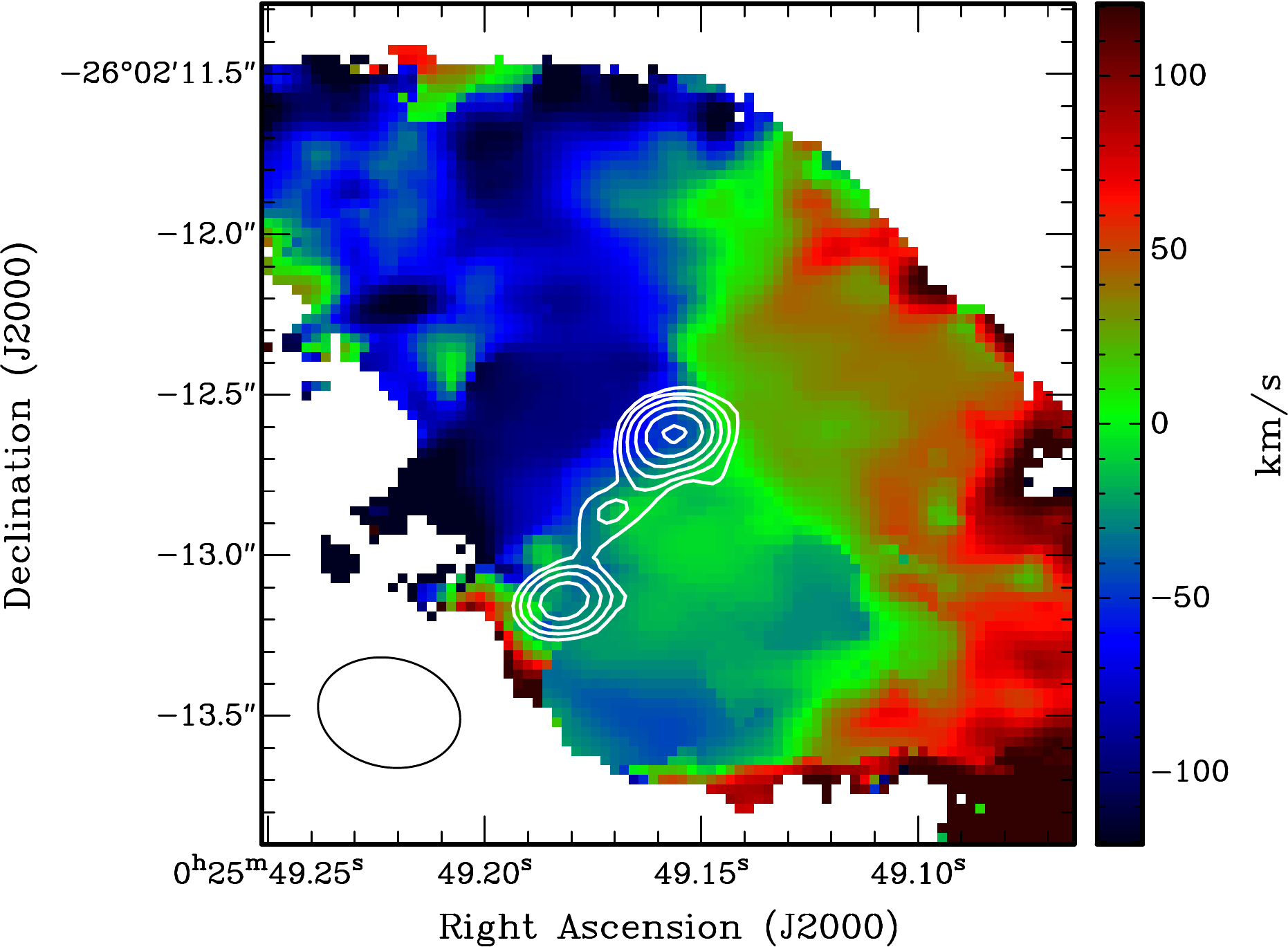}
      \includegraphics[angle=0,width=9cm]{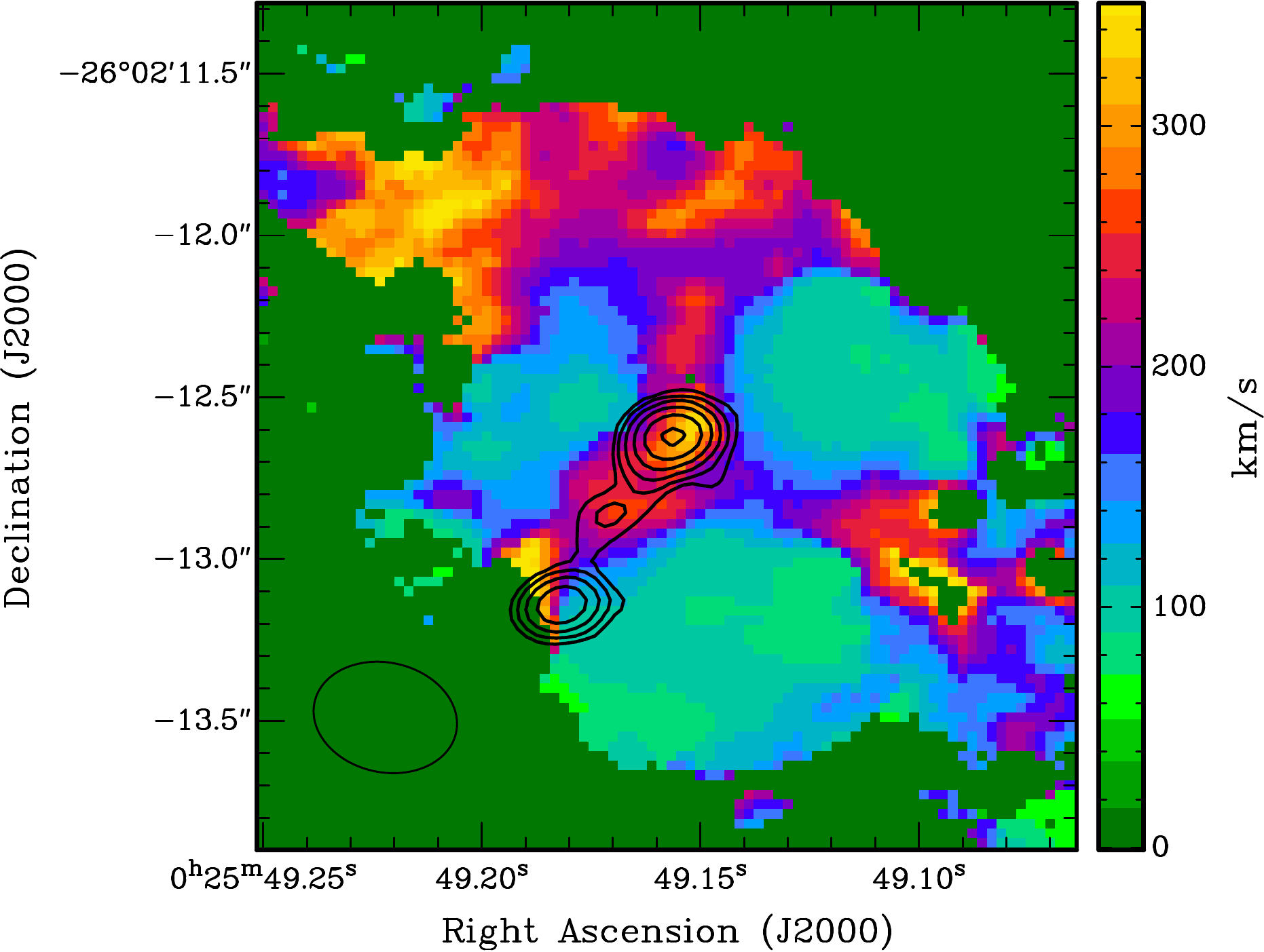}
   \caption{{\bf Left} Velocity field of the \coTwo\ with superposed the contours of the radio continuum emission. The overall velocity show a smooth gradient throughout the emission, although it is clear that the kinematics of the gas is complex. Contour levels are  1.3, 2.6, 5.2, 10.4, 20.8, and 41.6 \mJybeam. {\bf Right} Line width (FWHM) in colour with superposed the contours of the radio continuum emission. Contour levels are 1.3, 2.6, 5.2, 10.4, 20.8, and 41.6 \mJybeam. A region of high FWHM of the \coTwo\ line is clearly seen co-spatial with the radio continuum emission. In the northern region, the high velocity dispersion is the result of multiple stream of gas and multiple components of the velocity profile as can be seen in Fig.\  \ref{fig:profiles} region D.   
  }
              \label{fig:velocityField} 
    \end{figure*}

The total intensity image (Fig.\ \ref{fig:molecularGas}) was obtained from the naturally weighted cube in the standard way by smoothing this cube to twice the resolution and using the 2-$\sigma$ level of the smoothed cube to create a mask which was applied to the original cube. This image was converted to a column density image using a standard conversion factor $\alpha_{\rm CO} = 4.3 \ M_\odot/({\rm K\ km\ s^{-1}\ pc^2})$  \citep{Bolatto13} and assuming that $S_{\rm CO(2-1)} =   2.4\,S_{\rm CO(1-0)}$. The faintest emission detected in Fig.\ \ref{fig:molecularGas} has a column density of $10^{21}$ cm$^{-2}$. The velocity field and the  line width (FWHM) images (Fig.\ \ref{fig:velocityField}) were obtained using Gaussian fits to the line profiles for those pixels for which the line integral is larger than 10 mJy \kms.

Continuum images  were made using natural, uniform and intermediate (robust = 0.5) weighting.  The parameters of these images are summarised in Table \ref{tab:obs}.

%-------------------------------------- Continuum
%--------------------------------------

%-------------------------------------- molecular Gas
%--------------------------------------

%-------------------------------------- Velocity Field
%--------------------------------------
%-------------------------------------- Channels
   \begin{figure*}
   \centering
      \includegraphics[angle=-90,width=15cm]{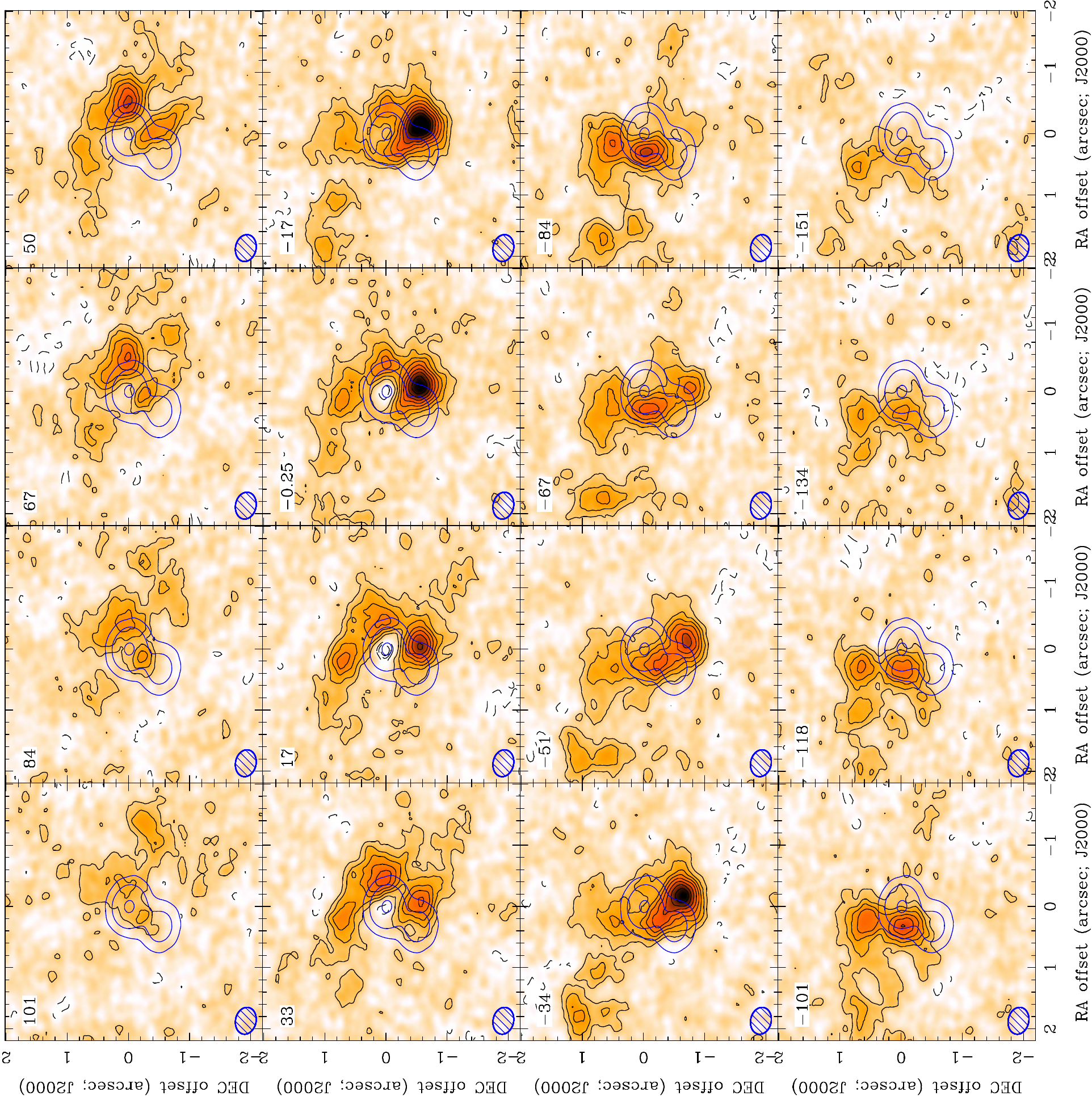}
   \caption{Channel maps of the \coTwo\ emission (from the natural weighting cube) in orange and black contours superposed to the  continuum emission (at the same spatial resolution) in purple. Contour levels for  the line emission are $ -0.36,$ 0.36 to 2.64 \mJybeam\ in steps of 0.24 \mJybeam\ and  4, 20, and 100 \mJybeam\ for the continuum. These plots illustrate how the gas is seen preferentially located around the lobes. In the central location, the emission is seen in multiple plots, confirming the  high velocity dispersion (as shown in Fig.\ \ref{fig:PositionVelocity}).
  }
              \label{fig:channels}
    \end{figure*}
%--------------------------------------

\section{Results: The 1.7 mm continuum emission}
\label{sec:continuum}

Because of the importance for relating the distribution of the gas to the radio source, we first describe the results for the 1.7 mm (176~GHz)  continuum emission. The  highest-resolution image (uniform weighting) is shown in Fig.\ \ref{fig:ContUnif}.
The same continuum emission is shown in contours  superposed to the column density image of the \coTwo\ in Fig.\ \ref{fig:molecularGas}, the velocity field and the line width images (Fig.\ \ref{fig:velocityField}). 

The structure of the continuum emission is dominated by two lobes, consistent with what is found at 2.3 and 8.4~GHz using  VLBI and at 5~GHz using MERLIN \citep{Tzioumis02}.  At the spatial resolution of our observations, the two lobes look quite symmetric. This is not the case at the  higher resolution of the VLBI observations, where the NW lobe shows a compact hot-spot, while the SE lobe has a more diffuse and resolved structure even on milli-arcsecond scales.  

The total flux measured across the broad band of the continuum observations (see Sect.\ \ref{sec:observations}) ranges from 87.7 mJy at 176.3 GHz to 96.1 mJy at 164.3 GHz. The fluxes of the two lobes at 176.3 GHz are listed in Table \ref{tab:fluxes} together with the  spectral indices obtained using the  fluxes at  2.3 GHz from \cite{Tzioumis02} and the 176 GHz fluxes of this paper. The total fluxes have been 
derived by fitting a 2-D Gaussian to each of the lobes.  For the fluxes derived from our ALMA data, we assume an error of 5\% while for the 2.3 GHz fluxes we assume an error of 10\%. The spectral indices we find over the large frequency interval between 2.3 and 176  GHz are quite normal for radio galaxies. However, because of the differences in spatial resolution, these spectral indices must be taken with care. It is worth noting that, while the spectral index for the northern lobe is consistent with what was found between 2.3 and 8.6 GHz, for the southern lobe the integrated index between 2.3 and 176 GHz is not as steep as derived in \cite{Tzioumis02} for the range 2.3 and 8.6 GHz. This is very likely due to missing flux in the 8.6 GHz VLBI images. 

\begin{table} 
\caption{Continuum flux densities and spectral indices derived from the ALMA and VLBI  observations. } 
\begin{center}
\begin{tabular}{cccccc} 
\hline\hline 
   & $S_{\rm 2.3\ GHz}$ &  $S_{\rm 176\ GHz}$ & $\alpha^{2.3}_{176}$\\
     &  (Jy) &  (Jy)\\
\hline
 NW - peak   & 1.07  & $0.048$ &  --0.72\\
 NW - total  & 1.60  & $0.059$ &  --0.76\\
 SE - peak   & 0.33  & $0.018$ &  --0.67\\
 SE - total  & 1.30  & $0.023$ &  --0.93\\
\hline
\end{tabular}
\tablefoot{The values at 2.3\ GHz are taken from \cite{Tzioumis02} and  we assume they have an error of 10\%.  The errors in the 176 GHz fluxes are dominated by the uncertainty in the flux scale, which we assume to be 5\%. The error in the spectral indices is 0.06.}
\end{center}
\label{tab:fluxes}
\end{table}

\section{Results: The molecular gas}
\label{sec:molecular}

We detect a  very extended distribution of molecular gas. This was not entirely unexpected based on the large IR luminosity (\citealt{Dicken09}, see also below). However, the distribution and kinematics of the molecular gas show some unexpected properties which are discussed in the following sections.

\subsection{The complex distribution of the molecular gas }
\label{sec:COdistribution}

The total intensity image of the molecular gas is presented in Fig.\ \ref{fig:molecularGas},  which also shows the location and extent of the continuum emission. Thanks to the relatively high spatial resolution of the observations, we can trace the gas along and around the radio source.

Most of the molecular gas is found in an extended structure in the galaxy of \pks. This structure has a  maximum extent  of $\sim$3 arcseconds (14 kpc) and is much larger than the radio continuum source. In addition, some clouds are detected at larger radii out to a  radius of  $\sim$5 arcseconds (24 kpc) from the nucleus.  The maximum recoverable scale in a single channel is 3.8 arcsec, but considering the  depth and high spatial resolution of the observations, we cannot exclude the possibility of missing  low surface brightness, diffuse emission. 
This is also suggested by the detection of some isolated regions at larger radii, possibly representing the tips-of-the-iceberg of a more extended structure of molecular gas.The overall distribution of the molecular gas is highly asymmetric, offset from  the centre of the radio galaxy and shows tails and filaments at larger radii. 

The brightest region of \coTwo\ emission is observed to coincide with the centre of the radio source and is extended in a direction   roughly perpendicular to the radio axis, and is $\sim 0.3$ arcsec (about 1 kpc) in diameter. The high surface brightness could be due to a relatively large amount of molecular gas piled up in this region, or to the presence of gas with higher excitation, as seen in other AGN, such as IC~5063 and PKS~1549--79 \citep{Oosterloo17,Oosterloo19}. Observations of other CO transitions are needed to clarify this. 

Interestingly, outside this bright central region, the molecular gas appears to show dips in brightness at the locations (in projection) of the radio lobes, especially at low velocities relative to the galaxy rest frame.  At these locations, the distribution of the molecular gas appears to ``embrace'' the radio emission (see Fig.\  \ref{fig:molecularGas} left). This is made clearer by the kinematics (see next section) and is relevant to our overall interpretation, as discussed in detail Sect.\ \ref{sec:COkinematics}.
Molecular gas in absorption is also detected against the peak of the continuum of the northern lobe (see Sect.\ \ref{sec:absorption}).

At  distances larger than about 1 arcsecond from the centre, the molecular gas is distributed in arm-like structures and clouds. From comparing these with the complex environment of \pks\ (Fig.\  \ref{fig:molecularGas}, right), it is clear that  one of these arms  points to an optical object located at about a 10 kpc projected distance to the east of the nucleus  (Fig.\ \ref{fig:environment}). If a separate galaxy, it may be the donor of at least part of the molecular gas. Alternatively, this stellar light may represent a tidal feature, part of the remnant of a much larger galaxy that is merging with \pks. Another arm or tail points to a similar optical feature at a similar distance from the nucleus on the western side.  
The distribution of gas also appears to be affected by the larger galaxy NE of \pks, which is  known to have have redshift similar to \pks. Indeed, some emission is observed at the location of C1. 
The possible origin of the large amount of molecular gas will be further discussed in Sect.\ \ref{sec:GasOrigin}.

%-------------------------------------- Velocity Dispersion
   \begin{figure}
   \centering
   \includegraphics[angle=0,width=8cm]{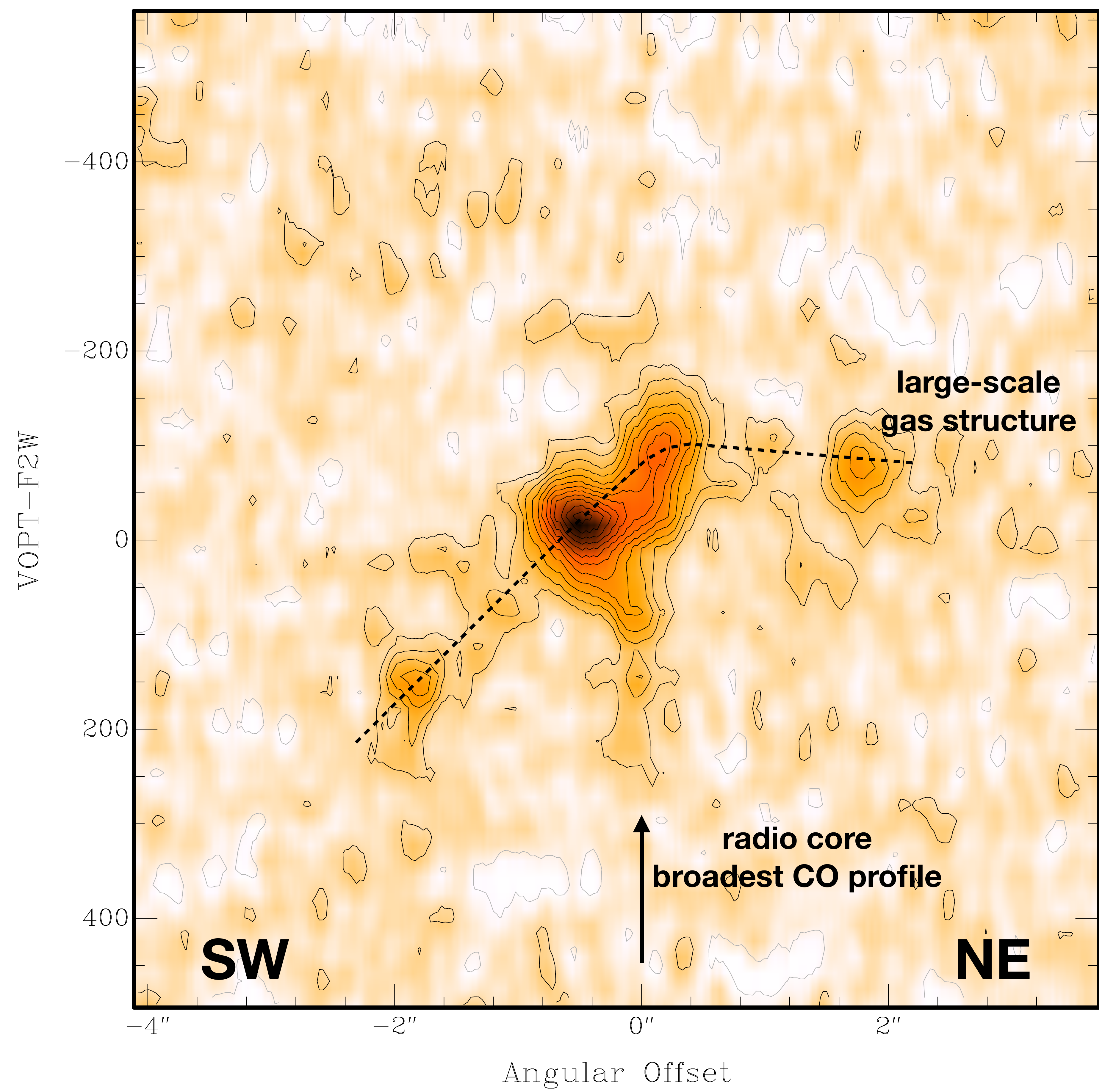}
   \caption{Position-velocity plot of the  \coTwo\ obtained from a cut 
   taken perpendicular to the radio axis and passing through the core (PA $= 52^{\circ}$). On the larger scales, the emission shows the overall velocity gradient of the CO and shows the  high velocity dispersion gas redshifted compared to the systemic velocity at the location of the core as a separate component.  Contour levels: $-0.18, 0.18$ to 2 \mJybeam\ in steps of 0.18 \mJybeam.
  }
              \label{fig:PositionVelocity}
    \end{figure}
%--------------------------------------

\subsection{The complex kinematics from small to large scales}
\label{sec:COkinematics}

\begin{figure*}
   \centering
          \includegraphics[angle=0,width=18cm]{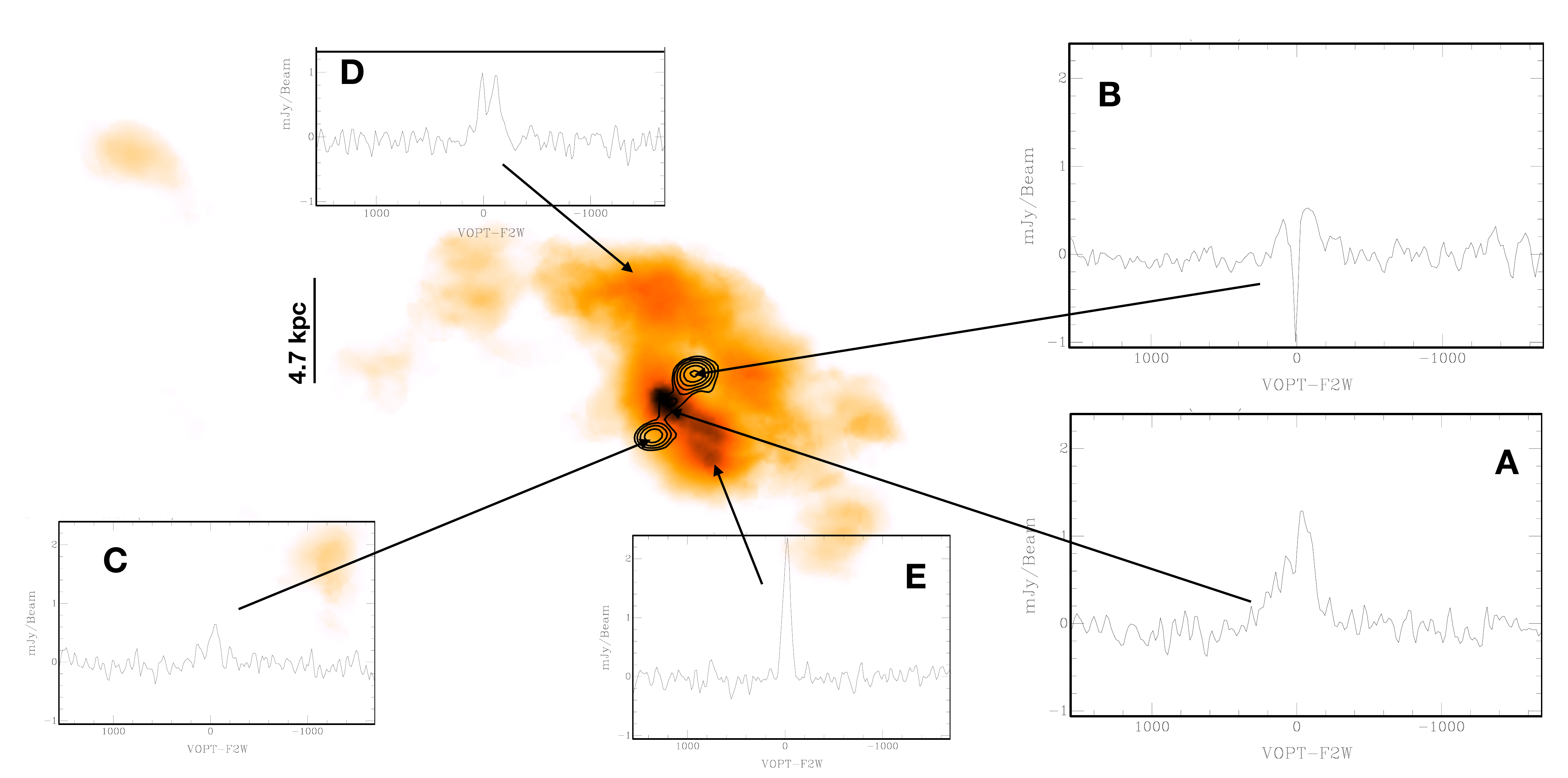}
   \caption{Profiles of the \coTwo\ taken from the low-resolution cube in various positions across the distribution of the molecular gas (in orange) and   the radio continuum (black contours): central region (A); peak of the radio lobes (B,C); the extended region of molecular gas away from the radio source (E) showing in some cases double profiles (D). The column density greyscale  is based on the same image as shown in Fig.\ \ref{fig:molecularGas}}
              \label{fig:profiles}
    \end{figure*}

The velocity field of the molecular gas is shown in Fig.\  \ref{fig:velocityField}, left. To first-order, a relatively smooth velocity gradient is seen throughout the distribution of the gas. However, the velocity field does not do justice to the complex 
velocity structure of the gas seen at various locations. Because of this, we show plots of the FWHM line width (Fig.\  \ref{fig:velocityField}, right), the channel maps (Fig.\  \ref{fig:channels}), position velocity plots (Fig.\  \ref{fig:PositionVelocity}) and line profiles (Fig.\  \ref{fig:profiles}) at key locations. In all the plots, the velocities are de-redshifted using the redshift, $z=0.32188$, derived by \cite{Santoro20}.

%-------------------------------------- Profiles plots
%--------------------------------------

In the bright central kiloparsec region,  between the two radio lobes and at the location of the radio core, we find that the width of the \coTwo\ profile is broader than in the other regions around the radio source. The profile at this location reaches FWZI $\sim$500 \kms\ and shows an asymmetric and redshifted wing (see profile marked (A) in Fig.\  \ref{fig:profiles}). 
The presence of this component is even better highlighted by the position-velocity plots in Fig.\ \ref{fig:PositionVelocity}. This slice is taken in a direction almost perpendicular to the radio axis. The channel maps in  Fig.\  \ref{fig:channels} show that, in this direction, a filament of gas is seen extending from NE to SW crossing the  location of the radio core. 
The position-velocity plot in Fig.\  \ref{fig:PositionVelocity}  shows a velocity gradient of the gas that may indicate the presence of a rotating structure or disc. However, and most interesting, this plot makes clear how the redshifted wing at the location of the radio core traces a separate component of unsettled gas deviating from the smooth kinematics of the rest of the larger-scale structure of the gas. 
Being in emission, it is difficult to say whether any of the gas with the redshifted velocities could be connected with feeding process (see Fig.\  \ref{fig:PositionVelocity} right).  

Moving outside the core region, the channel maps illustrate the distribution of the gas at the location of the radio lobes. As mentioned above, there is an absence of low velocity molecular gas at the peaks of the radio emission, giving the appearance that the gas wraps around the lobes, although the presence of the narrow absorption at the location of the N lobe may partly contribute   to this.  This can be seen better in the channel maps (see  Fig.\ \ref{fig:channels}, e.g., the velocity range 0 to 67\kms\ for the northern lobe) and is  more evident for the northern lobe. This indicates that the gas near the lobes is distributed in bubble-like structures.

The kinematics of the gas at the location of the lobes is not quiescent -- although the profiles are not as broad as those at the location of the core. This is well illustrated by the line-width map (see Fig.\  \ref{fig:velocityField}, right) where it can be noted that the region of large FWHM of the lines is well aligned and, in projection, co-spatial with the radio emission and that large widths are seen in particular at the lobes. 
As seen by the line-width plot, the emission profiles at the location (in projection) of the northern lobe are relatively broad (with a FWZI$\sim$350 \kms, see profiles marked with (B) and (C) in Fig.\  \ref{fig:profiles}), and symmetric around the systemic velocity. 
The width of the profile at the location of the southern radio lobe is narrower (FWHM $\sim$ 200 \kms). 

At larger radii,  well outside the region of the radio continuum emission, the gas shows a smooth overall  velocity gradient and the velocity dispersion is lower than near the lobes, for example region E in Fig.\ \ref{fig:profiles}. However, in the NE part of the large-scale CO distribution, as well as to the SW, the line widths are also quite large (Fig.\ \ref{fig:velocityField}). This is because here the line profiles show double peaks (e.g., (D) in Fig.\ \ref{fig:profiles}). These are likely connected with the tidal streams of gas pointing to the companions  of \pks\  (see also Sect.\  \ref{sec:GasOrigin}).

\subsection{Absorption against the NW radio lobe}
\label{sec:absorption}

\begin{figure}
   \centering
      \includegraphics[angle=0,width=7.75cm]{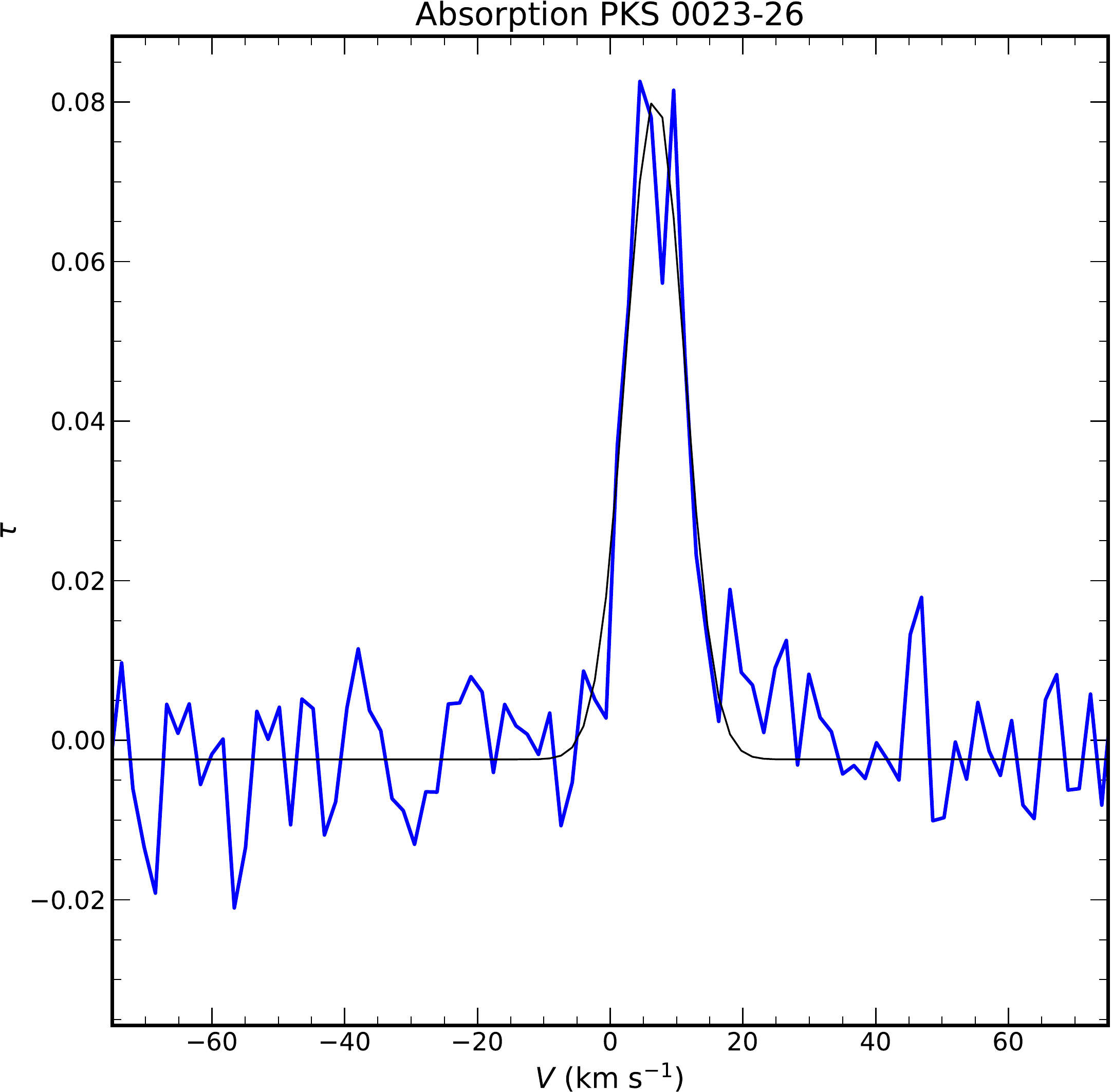}
   \caption{Absorption profile (blue) in units of $\tau$ at the location of the peak continuum of the NW lobe obtained from the highest spatial  (i.e., where most of the emission is resolved out) and spectral resolution cube. The black line is a Gaussian 
fit to the profile.
  }
              \label{fig:absorption}
    \end{figure}

We detect \coTwo\ in absorption at the location of the peak of the continuum emission of the northern lobe. Figure \ref{fig:absorption} shows the absorption profile  - plotted in units of optical depth $\tau$ - extracted, given the small velocity width, from the high spatial- and velocity resolution cube.   
The FWHM of the absorption is 23.0 \kms\ roughly centred on the systemic velocity. 
The depth of the absorption is 2.5 mJy against the continuum peak flux of 51 mJy. This corresponds to a peak optical depth of $\tau_{\rm peak} = 0.048$ and integral optical depth $\int \tau dv = 1.18$ \kms.  

The column density of the absorption can be estimated following the standard formulae such as those presented in  \cite{Bolatto13}. One of the assumptions needed is  the excitation temperature ($T_{\rm ex}$) of the gas. The typical temperature of the CO gas in conditions of thermal equilibrium in the Milky Way is $\sim$16 K (e.g., \citealt{Heyer09}). This temperature can be higher if the gas is affected by the AGN and this would increase the column density. Temperatures up to $T_{\rm ex}$ $\sim$60 K have been reported in some cases (see \citealt{Dasyra14,Oosterloo17}). The H$_2$ column density derived from the absorption in \pks\ ranges from $6.1 \times 10^{19}$ to $5.3 \times 10^{20}$ cm$^{-2}$ for  excitation temperatures of 16 and 60 K respectively, and a conversion factor CO-to-H$_2 = 10^{-4}$.
The narrow width of the absorption suggests that it is likely observed against a small, compact component in the NW lobe. Indeed, the high-resolution 8.3-GHz VLBI image presented in \cite{Tzioumis02} shows that this lobe hosts a compact hot-spot, with size $40\times 23$ mas (corresponding to $188 \times 108$ pc).
The width of the profile and the size of the likely background screen suggest that the absorption could be produced by a single giant molecular cloud (see \citealt{Miura21}). 

The column density derived for the H$_2$ in absorption is relatively low for molecular gas in general. 
In particular, it is much lower than those which are derived for the emission from nearby regions and which are at the level of $\sim$10 $\times 10^{22}$ cm$^{-2}$ (Sect.\ \ref{sec:mass}). One possible reason would be  that the absorption is partly filled by emission, reducing the amplitude. However, by looking at the profile in Fig.\  \ref{fig:absorption} at high spatial resolution, all emission seems to be  resolved out. A filling factor smaller than 1 could affect the column density derived for the absorption. However, to account for the difference, clouds much smaller than the typical size found in galactic ISM (see, e.g.,  \citealt{Miura21} for an overview and for the case of Centaurus~A) would be required to bring the two estimates of the column density closer together. Another possibility is that the molecular gas is quite clumpy on scales smaller than the beam and the absorption is caused by one of the lower  density clumps. This is supported by the fact that the profile from the lower-resolution data cube (marked with (B) in Fig.\ \ref{fig:profiles}) shows that the absorption has a narrower width than the emission at the location of the N lobe.
The more likely possibility is that the absorption and the bulk of the emission come from different locations, and that the absorption is due to a foreground cloud.

\begin{figure}
   \centering
      \includegraphics[angle=0,width=8cm]{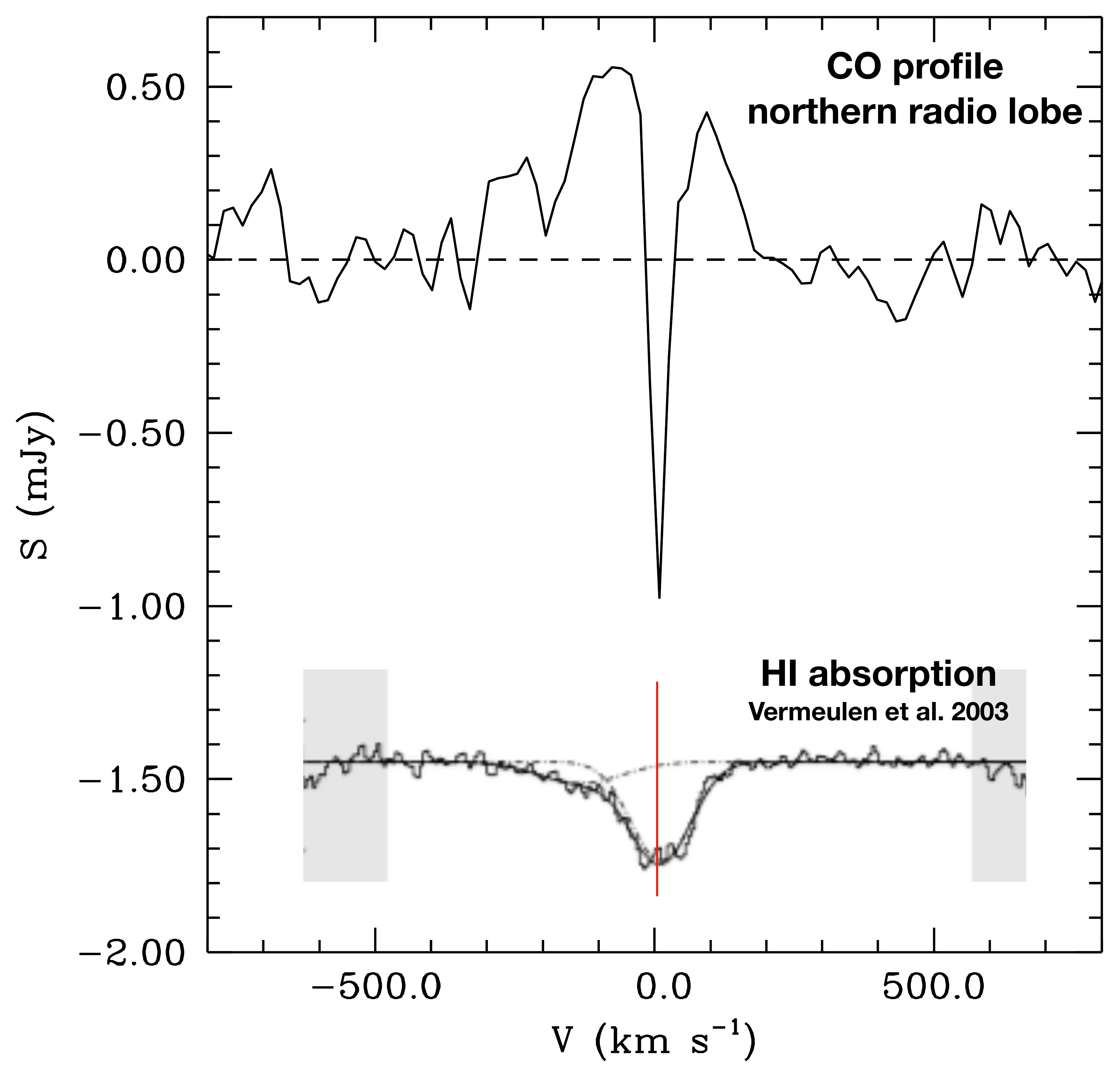}
   \caption{Comparison between the CO profile at the location of the northern lobe (from the lower-resolution cube) and the \HI\ absorption profile from \cite{Vermeulen03}. This illustrates the similarity of the velocity widths between the CO emission and the \HI\ absorption. The scales on the y-axis of the two plots are different. The regions marked in grey in the \HI\ plots were affected by radio frequency interference (see \cite{Vermeulen03} for the full discussion).
  }
  \label{fig:absorptionComparison}
  \end{figure}

Interestingly, the CO absorption profile is quite different from that of the absorption observed in \HI\ by \citet{Vermeulen03} (see  Fig.\ \ref{fig:absorptionComparison}). The \HI\ absorption is much broader and can be modelled by two components with FWHM = 126 and 39 \kms, respectively \citep{Vermeulen03}. The column density of the \HI\ was estimated to be $\sim 2.3 \times 10^{20}$ cm$^{-2}$ for $T_{\rm spin}=100$ K and a filling factor equal to 1. This value likely represents a lower limit to the \HI\ column density given that the filling factor is likely to be smaller than 1. 
In Fig.\ \ref{fig:absorptionComparison} a comparison is made between the CO profile at the location of the northern lobe (from the lower spatial resolution cube) and the \HI\ absorption profile from \cite{Vermeulen03}.  The difference between the two suggests that the \HI\ absorption does not trace the same gas as the CO absorption but, instead, traces at least part of the larger scale gas against the lobe, in a similar way as the CO emission does. This is illustrated by the similarity of the velocity widths of the CO emission and the \HI\ absorption. The difference between the absorption profiles of the two phases of the cold gas is likely due to differences in the radio continuum background between the relatively low frequencies of the 1.4~GHz \HI\ observations  (where the diffuse continuum emission of the lobe dominates) and the millimeter observations (where the emission of the hot-spot dominates).

%-------------------------------------- Absorption
%--------------------------------------

\subsection{Molecular masses}
\label{sec:mass}

The molecular mass was derived from the low-resolution cube.   The flux integral of the \coTwo\ emission is $S_{\rm CO(2-1)}\Delta v = 3.12 \pm 0.15$ Jy \kms, obtained by integrating the emission over the velocity range $-350$ to $+350$ \kms.
In order to derive the molecular mass, we first estimate the CO(1-0) luminosity $L^\prime_{\rm CO(1-0)}$ \citep{Solomon87,Solomon05,Bolatto13}:
\begin{equation}
L^\prime_{\rm CO(1-0)} =2453 \  D^2_{\rm L}\ (1+z)^{-1}\ S_{\rm CO(1-0)}\Delta\nu
\end{equation}
with $z$ the redshift, $D_{\rm L}$ is the luminosity distance in Mpc,  and $S_{\rm CO(1-0)}\Delta v$ is the integrated  flux in Jy \kms\ of the CO(1-0) line. 
In order to derive the $L^\prime_{\rm CO(1-0)}$, we have converted the \coTwo\ integrated flux to that of the ground transition \coOne\ by adopting two possible values for line brightness temperature ratio, $r_{21} = L^\prime_{\rm CO(2-1)}/L^\prime_{\rm CO(1-0)}=0.6$ and  $r_{21} = 1$. The former value is more consistent with quiescent gas, while the latter for disturbed gas affected by an AGN. This range is based on our results for IC~5063 and PKS~1549--79 \citep{Oosterloo17,Oosterloo19} and is also supported by  results on galaxy clusters \citep{Russell19} and high-redshift galaxies \citep{Aravena19}. 
The molecular mass was estimated using $M_{\rm H_2}=\alpha_{\rm CO}  L^\prime_{\rm CO(1-0)}$ 
where $\alpha_{\rm CO}$ is the CO-to-H$_2$ conversion  factor in units of \msun (K \kms\ pc$^2$)$^{-1}$.
We estimate the mass for two assumptions for $\alpha_{\rm CO}$: one is the value found for the MW and assumed for quiescent gas, $\alpha_{\rm CO} = 4.3$ \msun (K \kms\ pc$^2$)$^{-1}$. The other is 
$\alpha_{\rm CO} = 0.89$ \msun (K \kms\ pc$^2$)$^{-1}$, which is more typical for starbursts and ULIRG, (e.g.,  \citealt{Solomon87,Solomon05,Bolatto13}). 
With the assumptions described above, our estimates for the total molecular mass range from  $M_{\rm mol} = 3.1 \times 10^{10}$ \msun\ for a MW-like conversion factor and $r_{21}=1$,  to $M_{\rm mol} = 3.7 \times 10^{9}$ \msun\ for a conversion factor typical of ULIRGs and $r_{21}=0.6$. 
It is interesting to note that, using the dust mass of 1 -- 2.5 $\times 10^8$ \msun\ estimated by \citet{Bernhard21} and  assuming a standard dust-to-gas ratio of 100, this would imply a total (likely mostly molecular) gas mass of 1 - 2.5 $\times 10^{10}$ \msun, consistent with our estimates from the CO observations. 

%-------------------------------------- Comparison Emission-Absorption
%--------------------------------------

The  different  conditions of the molecular gas in  the nuclear regions close to the AGN compared to that of, for example, the gas at large radius and the  extended tails, likely require different conversion factors to be assumed for different regions. This was also found to be the case for, for example, IC~5063 and PKS~1549--79 \citep{Oosterloo17,Oosterloo19} based on CO line ratios, with a MW-like conversion factor required for the large-scale, regularly rotating gas and tails, and a lower conversion factor for the gas affected by the AGN. With the available data we cannot make a similar separation for \pks\ since observations of at least one other CO transition would be needed.

%================================================================
\section{Discussion: AGN feedback and evolution of the radio jets}
\label{sec:scenario}

The ALMA observations we present here show that the radio source \pks\ is embedded in an ISM rich in molecular gas.
The view obtained is complex  and, for some aspects, unexpected.
Below we attempt to put all  findings in the context of AGN feedback, jet-ISM interactions,  and the future evolution of the radio sources.

%-------------------------------------- Comparison Emission-Absorption
   \begin{figure}
   \centering
      \includegraphics[angle=0,width=7cm]{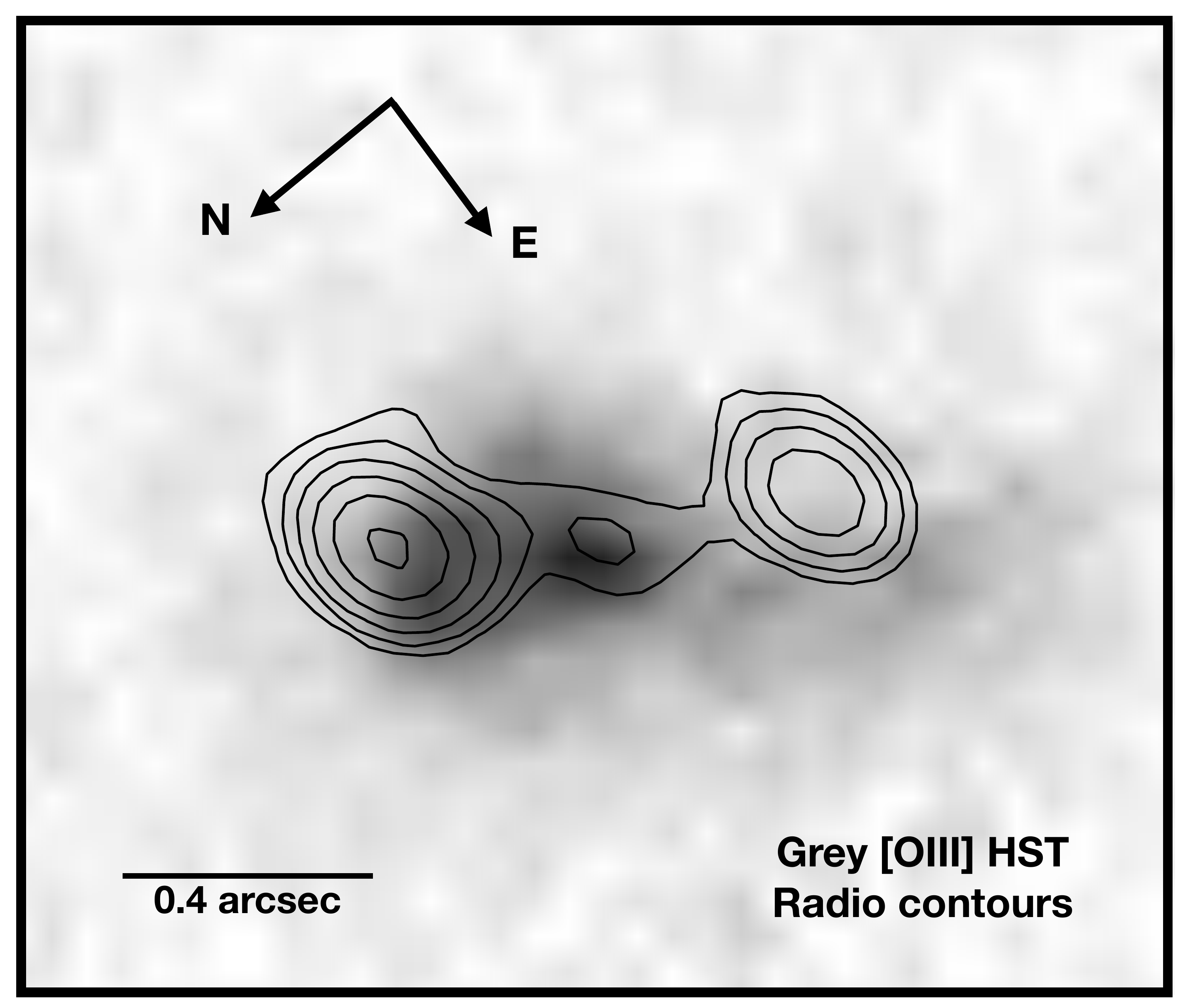}
   \caption{Overlay of the HST \OIII\ image from \citet{Santoro20} (grey scale) and the radio continuum contours from the high-resolution  continuum image.  Contour levels are as in Fig.\ \ref{fig:ContUnif}.
  }
              \label{fig:HSTradio}
    \end{figure}
%--------------------------------------

%================================================================

\subsection{Role of radiation and radio plasma jets}
\label{whichAGN}

\pks\ hosts both a radio and an optical AGN (see Sect.\ \ref{sec:description0023}). Therefore, it is important to consider which one plays the major role in affecting the kinematics of the gas. 
This question was also  addressed in the earlier studies of the ionised gas. The presence of gas with disturbed kinematics was already established from the finding of broad optical emission lines (FWHM $\sim 1000$ \kms) of warm ionised gas \citep{Holt08,Shih13,Santoro20}. The alignment between the radio axis and the distribution of the ionised gas was seen in the IFU data presented by \cite{Shih13}.

Further confirming this, Fig.\ \ref{fig:HSTradio}  shows the radio contours of the high-resolution continuum image overlaid on the \OIII\ emission observed with the HST. The latter image  was presented by \cite{Santoro20} but only now, thanks to the detection of the radio core, can we accurately align the two images. The \OIII\ emission is well-aligned with the radio axis, and is stronger towards the N lobe, although no brightening is observed at the centre of this radio lobe.
This alignment between the jet and the \OIII\ emission suggests that already from the warm ionised gas we can see the effect of the jet in shaping the distribution and kinematics of the gas. On the other hand, the ionisation of the gas is likely mostly due to photo-ionisation and not shocks  as discussed in detail by  \cite{Santoro20}, thus suggesting that only relatively mild shocks are present.  

As described in the introduction, \pks\ is a young radio galaxy. This is confirmed by the peaked radio spectrum as shown in \cite{Tzioumis02} and \cite{Callingham17}. The latter, by fitting  the radio SED, found the spectrum to peak at 145~MHz. 
If the peak is due to synchrotron self-absorption (SSA) - as suggested to be the case for most  radio sources in the young phase  -  the known correlation between size and peak frequency would result in an expected size of the source of $\sim 9$ kpc  \citep{ODea98}. While this value is uncertain given the large spread in the observed correlation, the smaller linear size of \pks\footnote{Given the relatively symmetric structure, the source is likely  close to the plane of the sky and no large correction for orientation to the size  is expected} may suggest that the expansion of the radio lobe has been slowed down due to the interaction with the rich medium. 
This suggests that the evolution of the radio source might have been affected (and slowed down) by the interaction with the surrounding  medium.

Thus, although we cannot completely rule out a role (especially in the inner kiloparsec) for radiation impacting the ISM, the alignment between the ionised gas and the radio axis as well as the disturbed kinematics along the radio source  (see below) suggest the radio plasma to be the most likely main mechanism affecting the gas kinematics.

\subsection{Impact of an expanding radio jet}
\label{sec:impact}

The ALMA results show molecular gas with relatively high velocity dispersion along the entire radio source, while a large amount of molecular gas with lower velocity dispersion extends to much larger radii. However, perhaps more relevant for the interpretation are the changes in kinematics and distribution of the gas that we see between the sub-kiloparsec and the kiloparsec scales. 

The most disturbed kinematics (with FWZI $\sim$ 500 \kms) of the molecular gas is found in the central region at the location of the radio core (well inside the inner kiloparsec). 
This is also the region where the molecular gas is brightest (Fig.\ \ref{fig:profiles}). The high brightness  could indicate a concentration of gas in the centre. However, in combination with the high velocities, it more likely indicate that the gas in this region has different excitation or optical thickness, as result of the impact from strong shocks.
Similar conditions have been found in other radio galaxies hosting molecular outflows  (i.e., IC~5063; \citealt{Oosterloo17}, and PKS~1549--79; \citealt{Oosterloo19}).  Observations of other CO transitions are needed to confirm this scenario in \pks.

%-------------------------------------- Cartoon ---------
   \begin{figure*}
   \centering
      \includegraphics[angle=0,width=14cm]{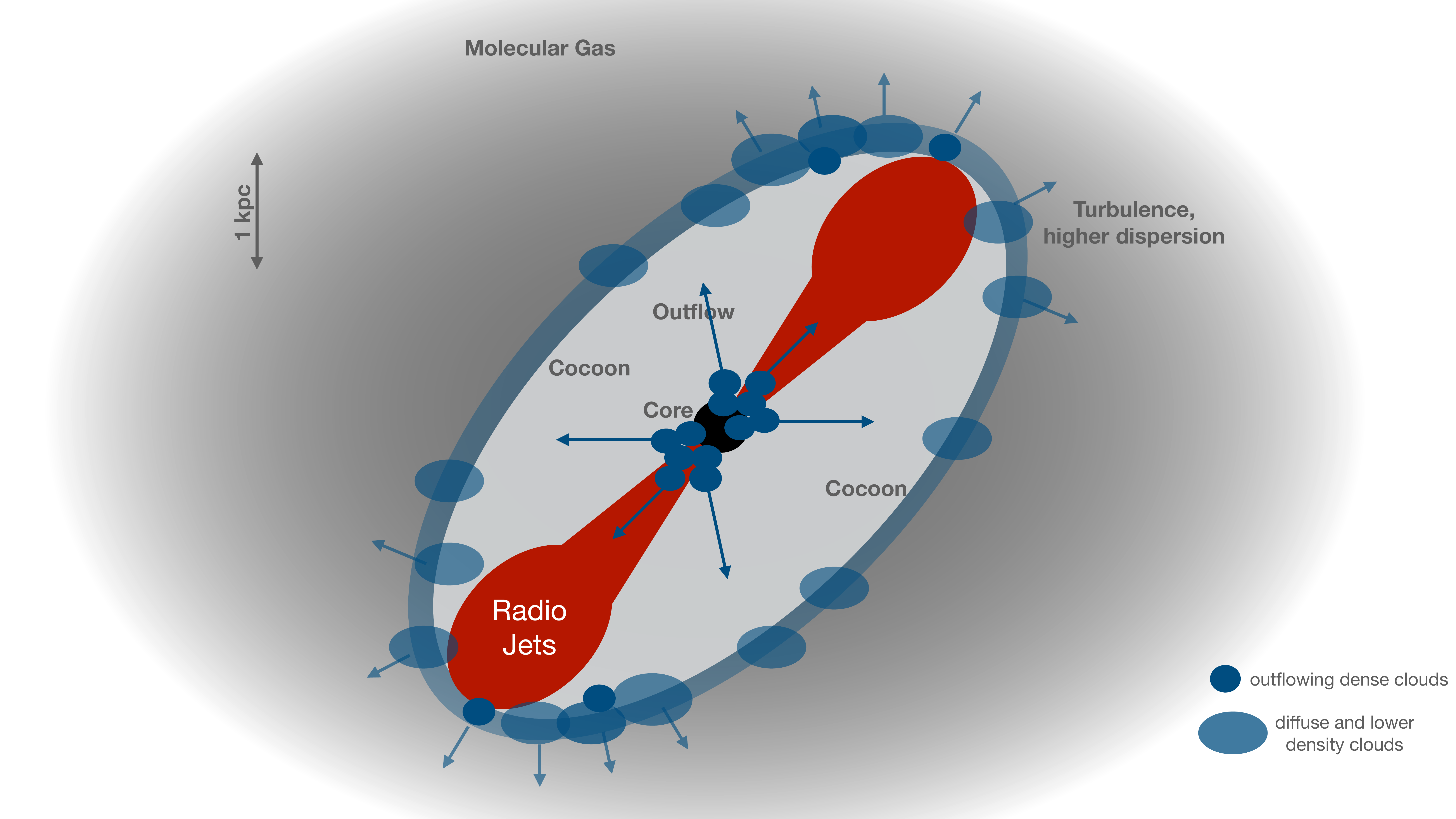}
   \caption{Cartoon illustrating the proposed scenario (see text for details). The length of the arrows indicates the higher velocities of the molecular gas observed in the central (sub-kiloparsec) regions compared to the larger scales. The dark grey area indicate the low velocity dispersion gas, the light grey indicates the regions - along the radio axis - where a dearth of molecular gas is seen. 
  }
              \label{fig:Cartoon}
    \end{figure*}
%--------------------------------------

The distribution of the gas outside the core region is remarkable: the  molecular gas with low velocity dispersion - distributed over many kiloparsec - tends to avoid the radio lobes, but broad lines - indicating larger turbulent velocities of the gas - are observed at  locations around the lobes (see Fig.\ \ref{fig:velocityField}). However, no strong outflows are detected. We consider these properties signatures of either the direct interaction of the radio lobes  with the ISM or, more likely given the not extreme velocities observed, the effect of the expanding cocoon of shocked gas created by the jet interacting with the clumpy ISM. Either way, the radio plasma is influencing and disturbing the molecular gas, not driving a significant outflow, but instead  dispersing and heating  preexisting molecular clouds,  excavating cavities devoid of dense, low velocity dispersion molecular gas.  

These findings appear to be  consistent with the predictions of the simulations. As described in the introduction, numerical simulations  (\citealt{Wagner12,Mukherjee16,Mukherjee18,Bicknell18} and refs therein) suggest that the jet starts with a phase of flood-and-channel, where it follows a meandering path of least resistance while strongly interacting with dense clouds in the ISM. The very broad profile observed in the  central (sub-kiloparsec) regions (Fig.\ \ref{fig:PositionVelocity}) suggests that this may be happening in this region. 
This phase of strong interaction also results in the creation of a cocoon of lower density shocked gas surrounding the jet, and an expanding energy-driven bubble, affecting a larger volume of the host galaxy, including in the direction perpendicular to the radio axis.
This would be the phase that produces the kiloparsec-scale features of the molecular gas around the radio lobes.

This scenario is summarised in the cartoon shown in Fig.\ \ref{fig:Cartoon}.
In the inner, sub-kiloparsec, region the main effect is the direct interaction between the newly born jet and the surrounding dense clouds. The presence of molecular gas in such extreme conditions has been observed in the inner regions of a number of radio galaxies (e.g., IC~5063, \citealt{Tadhunter14,Oosterloo17}; PKS~1549--26, \citealt{Oosterloo19}) and has been explained by the fast cooling of the dense gas (with pre-shock densities of a few hundred cm$^{-3}$ and  cooling times shorter than the dynamical times, see \citealt{Mukherjee18}). These conditions allow the reformation of molecules after the dissociation of the molecular gas resulting from the passage of the shock \citep[e.g.,][]{Tadhunter18}.

The simulations presented by \cite{Richings18}, although designed to describe the interaction of an AGN wind, show that, after being shocked, the dense gas can cool and (re)form molecules on very short timescales, within $\sim$1 Myr and will show up as a fast molecular outflow. Interestingly, these timescales are comparable with the time since the starting of the radio jet in \pks. 
If this is the origin of the outflow, it would also suggest that this phenomenon is a relative short phase in the life of the galaxy, although it can repeat multiple times as radio AGN are known to be recurrent \citep[see, e.g.,][]{Morganti17}.

At larger distances from the centre, the coupling between the radio plasma and the ISM is likely mediated by the cocoon of shocked gas created by the interaction jet-ISM, as predicted by the simulations. 
If this is the case, the non-extreme velocities we measure in the molecular gas at these locations might be due to the gas interacting with the (slower) expanding jet cocoon, rather than the faster head of the jet. This would be also consistent with the fact that the \OIII\ emission in the HST image avoids the head of the jet (see Fig.\ \ref{fig:HSTradio}).
Given the extended distribution of gas well outside the radio emission, the turbulent piling up around the radio lobes is likely the result of the expansion of the cocoon into the preexisting molecular gas, as sketched in Fig.\ \ref{fig:Cartoon}.
The energetics of the radio plasma turn out to be enough to do the pushing and piling up of the gas (see below).

The velocities observed at these locations are not extreme (and lower than those produced by the direct interaction of the jet with ISM clouds observed in the central regions), suggesting a relatively mild interaction resulting from the expansion of the cocoon. 
Whether the resulting physical conditions of the gas at this location are different (e.g., excitation temperature, more diffuse, lower density clouds etc.)  compared to those in the centre will need to be explored using the combination with other CO transitions.

In summary, the results of the observations and the  proposed scenario suggest that the impact of the radio plasma is both in ejecting gas (from the inner, sub-kiloparsec region) as well as in gently redistributing, heating and/or increasing the turbulence of the  gas on kiloparsec scales. This suggests a complex and varied way in which the AGN can affect the ISM and that could be relevant for the simulations of AGN feedback.  Indeed, a scenario including a similar complexity, with both ejective removal of  gas from the central kiloparsec and  preventative  heating of the circumgalactic gas, has been suggested by and is incorporated in the IllustrisTNG simulations for the case of wind and radiation feedback \citep{Zinger20}. 

The high jet power of \pks\ ensures  there is enough energy on kiloparsec-scales for accelerating and heating and dispersing the gas and forming the cocoon. The power required for creating the cocoon can be estimated by assuming that about $3 \times 10^9$ \msun\ of the molecular gas (i.e., about 10\% of the largest value derived for the total mass)  surrounds the radio lobes and that the gas is accelerated to a bulk velocity of (FWZI/2) $\sim$170 \kms\ and has a velocity width of $\sim$200 \kms, as discussed in Sect.\ \ref{sec:COkinematics}. In this case, the kinetic energy involved is on the order of a few times $10^{57}$ erg. Although this does not include the   energy to balance the gravitational potential of the host galaxy, the conservative assumptions made (e.g., the maximum value of the molecular mass involved) should provide a realistic estimate of the order of magnitude of the energy involved.  The value obtained is similar to that  found for some of the cool-core clusters, for example Abell~1795. 
 
In order to confirm that the radio jet can provide sufficient energy, we estimate  the jet power in \pks\ by using its low-frequency radio continuum flux.  With a radio luminosity of $\log L_{\rm 150~MHz}/{\rm erg\, s^{-1}} = 44.3$  (from a flux density at 145~MHz of 17.47 Jy from the GLEAM survey and reported in \citealt{Callingham17}), we estimate a jet power in the range  $3.5 - 5.9 \times 10^{46}$ \ergs\ using the relations presented in \citet{Willott99} and \citet{Cavagnolo10}.  Integrated over the life of the radio   source, assumed to be $10^6$ yr (quite typical for radio sources of this type), the total energy injected into the lobes over the lifetime of the source reaches $\sim$$10^{60}$ erg.
Thus, in contrast to what is often found for the radio galaxies in the centres of cool-core clusters (see, e.g., Abell~1795, \citealt{Russell17}), in \pks\ the jets appear to have enough energy to accelerate the gas and create the bubbles, even if the efficiency of the transfer of energy from jet to gas is low.

Finally, it is worth noting that \pks\ is not the only radio galaxy where gas structures, including molecular gas, have been seen to embrace the radio lobes. Similar structures have been found both in field- and cluster radio galaxies (see also Sect.\ \ref{sec:GasOrigin}). 
For example, extended molecular filaments are observed around radio bubbles in at least 6 of the 12 central galaxies in cool-core clusters summarised in \cite{Russell19}.
Among them, the two best examples of molecular gas ``hugging'' the radio lobes have been found in PKS~2322--123,  a young radio source at the centre of Abell~2597, and in 4C~26.42 at the centre of Abell~1795, by  \cite{Tremblay18} and  \cite{Russell17} respectively. 
However, given the lower jet power of these radio sources, the distribution of molecular gas could be the result of a different process compared to \pks. 

Bubbles of gas produced by the expansion of the radio lobes have also been seen  in galaxies outside clusters, albeit mostly in the X-ray and/or ionised gas. The list of objects includes  low-power radio sources (e.g., Circinus, \citealt{Mingo12}; Mrk~6, 
\citealt{Mingo11}; NGC~3801 \citealt{Croston07}; TeaCup, \citealt{Lansbury18}) as well as intermediate and powerful radio galaxies (Centaurus~A, \citealt{Croston09}; Coma~A and 3C~171, \citealt{Tadhunter00}; 3C~305, \citealt{Hardcastle12}; PKS~2250--41, \
\citealt{Villar99}; 3C273, \citealt{Husemann19b}). In Coma~A, the ionised gas wrapping around the radio lobe has been explained using shocks of 200 \kms, similar to what could be happening in \pks. Unfortunately, deep and high spatial resolution observations tracing the molecular gas are not available for these objects, preventing confirmation of whether an evolutionary sequence can be seen in the gas tracing the expansion of the radio lobes, and under which circumstance the molecular gas is still present.

%-------------------------------------- Plots from MS
   \begin{figure}
   \centering
      \includegraphics[angle=0,width=8.5cm]{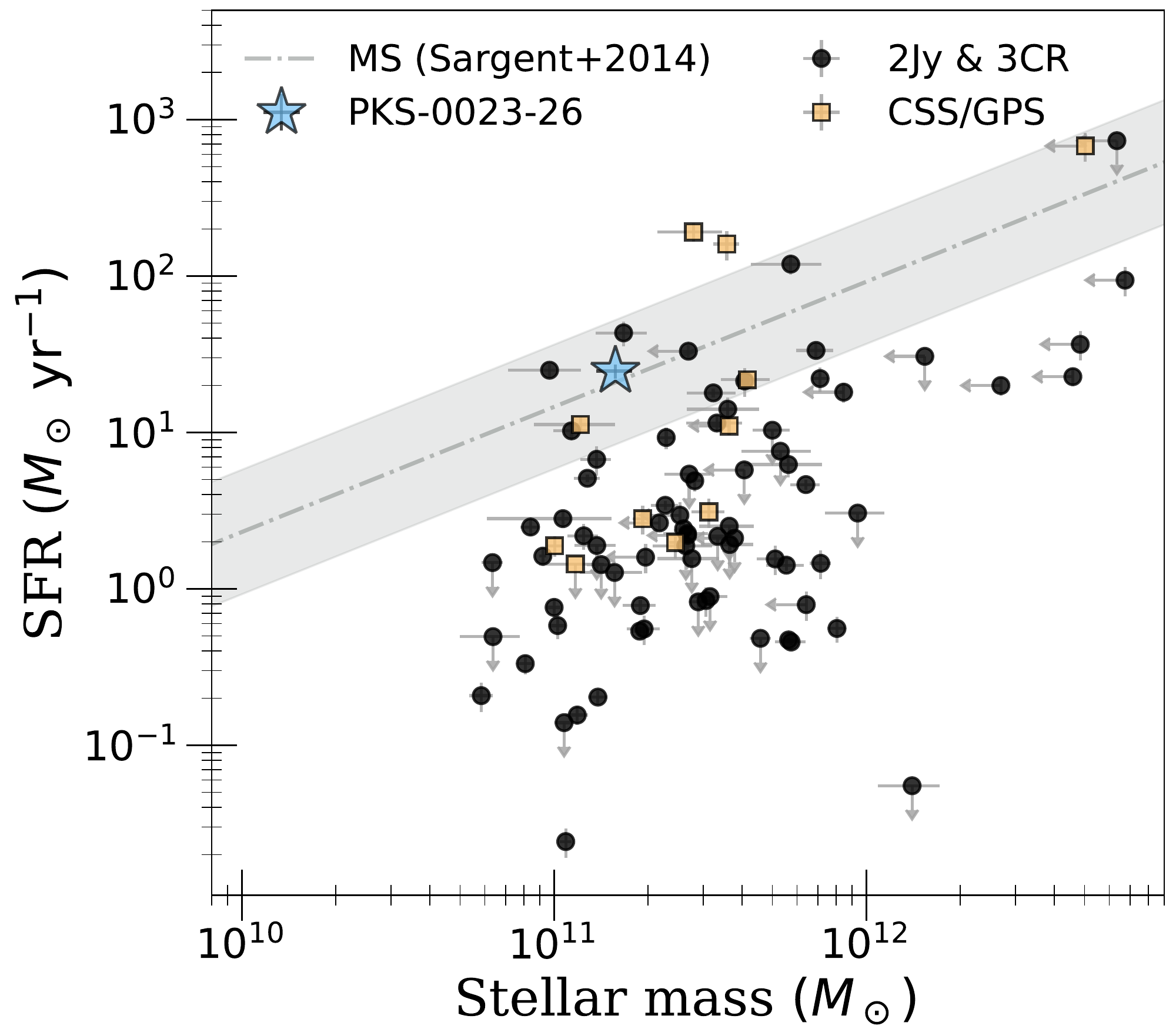}
   \caption{{Plot of the star-formation rate vs stellar mass for the 2~Jy and 3CR samples of radio galaxies at $z<0.7$ and $z<0.5$  respectively (from \citealt{Bernhard21}). The CSS and GPS sources are marked in yellow, and \pks\ highlighted as the lightblue star. The black line shows the star-formation main sequence (from \cite{Sargent14}) at the exact redshift of \pks, and the shaded area is the scatter of 0.4 dex.}
  }
              \label{fig:MS}
    \end{figure}
%--------------------------------------

\subsection{Gas, star formation and future evolution}
\label{sec:starformation}

As mentioned in the introduction, the host galaxy  shows a large population ($\sim$50\%) of young  stars (YSP) with ages estimated to be between 30 and 50 Myr \citep{Tadhunter11,Ramos13}. This is interesting in the context of the impact of the AGN on the host galaxy and the timescale of the interaction.
In fact, \pks\ has the largest fraction of YSP of the entire sample of radio galaxies studied by \cite{Dicken09}  (see also \citealt{Tadhunter16} for a review).
The region with young stars extends to at least 15 kpc (\citealt{Holt07}), suggesting a connection between the accretion event that brought in the molecular gas and the star formation (see Sect.\ \ref{sec:GasOrigin}). 
Like in other objects, the burst of star formation (with ages  3 -- 5 $\times\ 10^7$ yr)  appears to follow with some delay the merger (with an age of at least $10^8$ yr; see below), while the starting of the AGN radio emission ($\sim$10$^6$ yr) follows with an additional delay  (e.g., \citealt{Emonts08,Tadhunter11}).  

Interestingly, considering the SFR of  { $25\pm 2$ \msunyr\ and the stellar mass of the host galaxy of \pks\ { ($M_* = 1.6\pm 0.4 \times 10^{11}$ \msun;  \citealt{Bernhard21}}), this galaxy appears to be located on the SFR-$M_{*}$ relation for star forming galaxies.
Figure \ref{fig:MS} shows that, while \pks\ is one of the most highly star forming objects in the 2~Jy sample, it appears to lie on the main sequence for star forming galaxies at $z=0.3$ from \cite{Sargent14}; also its level of star formation is not unusual for powerful CSS and GPS sources with similar redshifts and host galaxy stellar masses.
This is also the case when the properties derived for \pks\ are compared with what is found for galaxies of similar stellar mass at low and high redshift (e.g., \citealt{Genzel10,Genzel13}). The molecular gas properties and gas depletion times of \pks\ are also consistent with the results from the ALMA Spectroscopic Survey in the Hubble Ultra Deep Field (ASPECS)  \citep{Aravena19}. 

A possible scenario of what is happening in \pks\ could be that, despite the powerful radio and optical AGN, in \pks, at this stage of the evolution  the effect of the AGN, and in particular of the radio AGN, is limited to the kiloparsec scales. The large amount of molecular gas distributed on larger scales present in the galaxy is not (yet) directly affected.  In the case of \pks, the associated large-scale star formation has not been quenched, at least not yet. However, as we have shown above, as the jet and lobes expand in the coming few $\times 10^7$ yr, they carry enough energy to prevent more gas from cooling, potentially regulating the star formation on lager scales. 
Given the observed properties of the gas at larger scales in \pks, the longer term impact on the ISM is likely more related to the relatively mild expansion of the cocoon and lobes to larger scales, and the consequent heating of the ISM and circumgalactic gas, instead of only the gas outflow.
Interestingly, this resembles the prescriptions  included in recent cosmological simulations \citep[e.g,.][]{Zinger20}, which, however, did not focus on radio AGN.

Finally, it is worth noting that the scenario in which star formation is not rapidly suppressed by an AGN  has been suggested by the results obtained by \cite{Scholtz18} and \cite{Harrison21} who have explored feedback models  using EAGLE cosmological simulations. They find that the feedback likely acts over longer timescales.
Our results bare also similarities with the scenario proposed by  \cite{Ellison21} using the EDGE-CALIFA sample and finding lower gas fraction in the central kiloparsec-scale region of AGN compared to that in star forming regions, suggesting an inside-out quenching. 
Thus, all these studies, although looking at different types of AGN, appear to point to a more complex view of AGN feedback.

\subsection{Origin of the molecular gas}
\label{sec:GasOrigin}

A large amount of molecular gas is detected in \pks, raising the question of its origin. This is also relevant for learning about the triggering of the AGN and   the origin of the large FIR luminosity,  whether they are the result of a major merger (as many powerful AGN are considered to be), or whether the accretion is due to phenomena connected to smaller interactions with galaxies from  the rich environment in which \pks\ is embedded.

The mass of the molecular gas (a few times the total gas mass of the Milky Way) suggests that the host galaxy of \pks\ is possibly the result of a major gas-rich merger, and that the large-scale structures represent the debris of this merger. The optical morphology and the presence of a common stellar envelope in which \pks\ and some of the neighbouring galaxies are embedded support the major merger scenario  (see Fig.\ \ref{fig:environment} and \citealt{Ramos13}). However, the presence of tidal streams suggests that at least part of the gas is being accreted  from   companions. The presence of double peaked profiles in the gas kinematics are indications for the superposition along the line-of-sight of different gas components that have not yet settled into a stable dynamical configuration (see the region marked  D in Fig.\ \ref{fig:profiles}). 

If part of the gas is coming from a merger or accretion process, the fact that the gas appears to be still unsettled can provide a rough estimate of the timescale of the process.  Considering the distribution and kinematics, the gas has not yet completed several orbits around \pks\ to build a settled disc. A rough estimate of the age of the merger would be to assume the gas has made   one orbit at the radius of 10 kpc which gives timescale of a few times $10^8$ yr. 

However, some of the observed properties are difficult to reconcile with (only) a merger scenario, the mass of the molecular gas observed in \pks\ is large compared to what is typically found in early-type galaxies in groups and clusters and most radio galaxies  have molecular gas masses ranging between a few $10^7$  \msun\ and a few $10^8$ \msun\ \citep{Young11,Ruffa19}.
\pks\ would also fall in the upper end of the range  for quasar-like AGN in the local Universe. For a sample of high-power radio galaxies  \citet{Tadhunter14} and \citet{Bernhard21} found,   using the dust masses, molecular gas masses  in the  range $10^8$ \msun\ to  $5\times10^{10}$\ \msun\ (with a mean value of $\sim$2$\times 10^9$ \msun). 
Furthermore, the molecular gas structure in \pks\ is much more extended than typical for early-type galaxies in groups and clusters \citep[e.g.,][]{Davis13}. Most of such galaxies have no or very little gas and if discs of molecular gas are observed in such galaxies, they typically have a size of a few kiloparsec, up to at most 10 kpc. Thus, the distribution of molecular gas in \pks\ seems to exceed what is usually found in early-type galaxies in dense environments. 

On the other hand, the properties of the molecular gas observed in \pks\ bear some similarities to those observed in galaxy clusters and, in particular, cool-core clusters.
For example, in cool-core clusters, the distribution of molecular gas is often offset from the centre, the gas is partly distributed in filaments (often wrapping around the radio emission) and shows a combination of smooth velocity gradients and regions of large velocity dispersion (see \citealt{Tremblay18} and \citealt{Russell19} for overviews). Regular discs of molecular gas are rare in clusters (see Hydra~A and A262 for two exceptions, \citealt{Russell19}). The amount of gas, the large size and the offset distribution of the molecular gas in \pks\ is reminiscent of this. 
As suggested by \cite{McNamara16,Russell19}, most of the molecular clouds in cool-core clusters may have formed from low entropy X-ray gas that became thermally unstable and cools as it is lifted by the buoyant radio bubbles.
However, the X-ray luminosity of \pks\ suggests a group environment rather than a rich cluster environment (see Sect.\  \ref{sec:description0023} and \citealt{Eckmiller11}). Although galaxy groups can be associated with cool cores \citep[e.g.,][]{OSullivan15,Werner14}, and molecular gas is sometimes detected in the central group galaxies \citep{OSullivan18,Schellenberger20}, the typical molecular gas masses are modest ($<10^9$ \msun) -- an order of magnitude lower than what we measure for \pks. On this basis, rather than having cooled from the hot X-ray halo, it seems more plausible that the molecular gas in \pks\ has been accreted from other galaxies. 

In conclusion, the origin of the molecular gas in and around the host galaxy of \pks\ is still unclear and requires a better understanding of its environment. Hopefully the deeper Chandra observations already planned will help clarifying the presence - or not - of a cluster environment.

%================================================================

\section{Summary and conclusions} 

Thanks to the combination of a rich ISM and a favourable geometry,  we have been able, using  high spatial resolution ALMA observations, to trace the details of the interaction between the radio-plasma and the molecular gas in the young, kiloparsec-scale radio galaxy \pks.

We detect a large amount of molecular gas (between 0.37 and $3.1 \times 10^{10}$\msun, depending on the assumptions of the conversion from CO to H$_2$), distributed over a radial extent up to $\sim$5 arcsec (24 kpc), and extending well beyond the radio source. 

The gas shows disturbed kinematics along the full extent of the radio source, a signature of the effects of the radio plasma on the ISM.
However, thanks to the high spatial resolution of the ALMA observations, in \pks\ we can see a change in the  way the radio plasma couples to the gas when going from the sub-kiloparsec to the kiloparsec regions.
The strong  direct interaction between radio plasma and gas appears to be limited (as in other radio galaxies) to the sub-kiloparsec scale. This results not only in the most disturbed kinematics (FWZI $\sim$500 \kms), but also in the peak of the intensity of the molecular gas. 
On kiloparsec-scales, the molecular gas tends instead to follow the edge of the radio emission with kinematics that do not reach extreme velocities (FWZI $< 350$ \kms). This suggests that at these locations  the relatively mild expansion of the cocoon, created by the interaction between jet and ISM, into the preexisting molecular gas is pushing the gas aside. 
This affects the gas in a  gentler way, resulting in dispersing and heating the preexisting molecular clouds. 

The observed velocities of the gas are not extreme, despite the presence of a powerful AGN and the large amount of energy (radiative and mechanical) released by the nucleus.
This result adds to the growing number of studies suggesting that AGN-driven gas outflows - even if very common in AGN - cannot account alone for the  feedback effects required by simulations of galaxy evolution. This is the case for objects  both at low 
as well as high redshift (see, e.g.,  \citealt{Bischetti19}).
Indeed, more subtle signatures are starting to be found (see, e.g.,  \citealt{Ellison21}), and there is the  possibility that the effects of AGN on the ISM are not instantaneous, but rather more long-term effects as the cocoon and lobes expand in a relatively mild way to larger radii  (see, e.g., \citealt{Scholtz20,Harrison21}). 

The properties of the molecular gas observed in and around the radio source \pks\ suggest that we are tracing different phases of the jet-ISM interaction and, therefore, different ways in which the energy is exchanged between the radio plasma and the ISM: from dominated by outflows in the inner region, to inducing turbulence by the cocoon created by the expansion of the radio jets.
Interestingly, the possibility of having both ejective (removing the central gas reservoir) and preventative (heating the circumgalactic gas) is now included in the implementation of AGN feedback in the IllustrisTNG cosmological simulation \citep{Zinger20}. Although only focused on the effects of winds and radiation (and not of the impact of radio sources), this paper still demonstrates the need for a more realistic implementation of the impact of the AGN.

Our results on \pks\  further highlight the relevance of galaxy-scale radio sources \citep[see also][]{Webster21}, which can provide the main exchange of energy between the radio plasma and the cool ISM, as well as a  bridge to the CGM and IGM, where large radio sources are known to have large impact. 

\begin{acknowledgements}
The authors would like to thank Beatriz Mingo and Aneta Siemiginowska for their help with the X-ray data and Luca Oosterloo for his help with some of the figures. This paper makes use of the following ALMA data: ADS/JAO.ALMA\#2018.1.01598.S. ALMA is a partnership of ESO (representing its member states), NSF (USA) and NINS (Japan), together with NRC (Canada), MOST and ASIAA (Taiwan), and KASI (Republic of Korea), in cooperation with the Republic of Chile. The Joint ALMA Observatory is operated by ESO, AUI/NRAO and NAOJ. CT and EB acknowledge support from STFC.
\end{acknowledgements}

\end{document}